\documentclass[iop,revtex4,apj]{emulateapj}

\usepackage{epstopdf}
\usepackage{natbib}
\usepackage{xspace}
\usepackage{amsmath}
\usepackage{lscape}
\newcommand{\Osix}{\ion{O}{6}\xspace}
\newcommand{\Ofour}{\ion{O}{4}\xspace}
\newcommand{\Othree}{\ion{O}{3}\xspace}
\newcommand{\Otwo}{\ion{O}{2}\xspace}
\newcommand{\Oone}{\ion{O}{1}\xspace}
\newcommand{\Cthree}{\ion{C}{3}\xspace}
\newcommand{\Ctwo}{\ion{C}{2}\xspace}
\newcommand{\Ntwo}{\ion{N}{2}\xspace}
\newcommand{\Nthree}{\ion{N}{3}\xspace}
\newcommand{\Nfour}{\ion{N}{4}\xspace}
\newcommand{\Nfive}{\ion{N}{5}\xspace}
\newcommand{\Hetwo}{\ion{He}{2}\xspace}
\newcommand{\Hone}{\ion{H}{1}\xspace}
\newcommand{\Neeight}{\ion{Ne}{8}\xspace}

\newcommand{\Sitwo}{\ion{Si}{2}\xspace}

\newcommand{\Cloudy}{\textsc{Cloudy}\xspace} 
\newcommand{\e}[1]{\ensuremath{\times 10^{#1}}}

\begin{document}

\title{Constraining UV Continuum Slopes of Active Galactic Nuclei With CLOUDY Models of Broad Line Region EUV Emission Lines \footnotemark[1]}
\footnotetext[1]{Based on observations made with the NASA/ESA {\it Hubble Space Telescope}, obtained from the data archive at the Space Telescope Science Institute. STScI is operated by the Association of Universities for  Research in Astronomy, Inc. under NASA contract NAS5-26555.}

\author{Joshua Moloney and J. Michael Shull\altaffilmark{a}}
\affil{CASA, Department of Astrophysical \& Planetary Sciences, \\University of Colorado, Boulder, CO 80309}
\email{joshua.moloney@colorado.edu,\linebreak michael.shull@colorado.edu}
\altaffiltext{a}{Also at Institute of Astronomy, University of Cambridge, Cambridge CB3 0HA, UK}

\begin{abstract}
Understanding the composition and structure of the broad-line region (BLR) of active galactic nuclei (AGN) is important for answering many outstanding questions in supermassive black hole evolution, galaxy evolution, and ionization of the intergalactic medium. We used single-epoch UV spectra from the Cosmic Origins Spectrograph (COS) on the \textit{Hubble Space Telescope} to measure EUV emission-line fluxes from four individual AGN with $0.49 \le z \le 0.64$, two AGN with $0.32 \le z \le 0.40$, and a composite of 159 AGN. With the \Cloudy photoionization code, we calculated emission-line fluxes from BLR clouds with a range of density, hydrogen ionizing flux and incident continuum spectral indices. The photoionization grids were fit to the observations using single-component and locally optimally emitting cloud (LOC) models. The LOC models provide good fits to the measured fluxes, while the single-component models do not. The UV spectral indices preferred by our LOC models are consistent with those measured from COS spectra. EUV emission lines such as \Nfour $\lambda$765, \Otwo $\lambda$833, and \Othree $\lambda$834 originate primarily from gas with electron temperatures between 37000~K and 55000~K. This gas is found in BLR clouds with high hydrogen densities ($n_H\ge10^{12}$ cm$^{-3}$) and hydrogen ionizing photon fluxes ($ \Phi_H \ge10^{22}$ cm$^{-2}$~s$^{-1}$).
\end{abstract}


\section{INTRODUCTION}
\setcounter{footnote}{1}

Supermassive black holes (SMBH) are thought to lie at the centers of almost all massive galaxies. Ever since the discovery of relationships between SMBH mass and host galaxy properties such as bulge mass \citep{kormendy93,kormendy95} and stellar velocity dispersion \citep{ferrarese00,gebhardt00,gultekin09}, understanding the mechanisms governing these relationships has been a major focus of galaxy evolution research \citep{kormendy13}. SMBHs are thought to grow primarily during periods of rapid accretion, during which the black hole and its surrounding gas are highly luminous, a phase referred to as a thermal active galactic nucleus (AGN). These periods may be triggered by changes in the host galaxy such as occur during a merger \citep{silk98,dimatteo05}. In turn, the AGN can influence its host through mechanisms such as large-scale outflows \citep[e.g.,][]{mcnamara05,arav08,dunn10}, which may enhance or inhibit star formation over large regions \citep[e.g.,][]{croton06,silk09}. The details of these mechanisms, along with the length of AGN duty cycles \citep[e.g.,][]{hopkins06,goncalves08}, are all active areas of research.

AGN also play an important role in \Hetwo reionization at redshifts $z\approx3$. Stellar populations do not produce sufficient radiation at energies above 4 Ryd to reionize \Hetwo, so AGN, with their harder spectra, are thought to be the sources primarily responsible for the \Hetwo reionization epoch \citep{reimers97,shull04}. Understanding the length of AGN duty cycles, thought to last $10^7 - 10^8$ years, along with the shape of the \Hetwo ionizing continuum, is of central importance for accurately modeling this period. There is evidence that the shape of the intrinsic AGN EUV continuum is harder than what we observe \citep{korista97_2}. If this is true, then understanding how and where the continuum changes will be important for \Hetwo reionization models.

For redshifts below $z\approx3$, AGN are predicted to dominate the metagalactic ionizing background \citep{haardt12}. This radiation plays a major role in determining the distribution of absorbers in the Ly$\alpha$ forest and the ionization states of metals in the intergalactic medium (IGM). Much of what we know about the distribution of baryons in the low-$z$ IGM comes from absorption line studies of Ly$\alpha$ \citep{penton00,penton04,lehner07,danforth08,tilton12} and metals such as \Osix \citep{danforth08,tripp08,thom08}. With over 75\% of the baryons at $z = 0$ potentially located in various thermal phases of the IGM, an accurate picture of its ionization state will be necessary to resolve the ``missing baryon'' problem \citep{shull12}. In the circumgalactic medium around galaxies hosting AGN, the effects of ionization are even more pronounced \citep[e.g.,][]{carswell87,bajtlik88,goncalves08} and may linger for tens of millions of years after the end of the AGN phase due to long metal recombination timescales \citep{oppenheimer13}. 

Only the nearest SMBHs can have their gravitational spheres of influence resolved directly, so almost all of what we know about the structure of AGN is inferred from their spectra \citep{baldwin97}. The high ionizing fluxes present in AGN lead to emission lines from a large number of ions. Because these lines are extremely broad, with widths of several thousand km s$^{-1}$, the region where they form is referred to as the broad-line region (BLR). Reverberation mapping \citep{blandford82, peterson93,vestergaard06} has led to many recent advances in our understanding of the structure of the BLR \citep[see][for a recent review]{gaskell09}. Reverberation mapping measures the response times of emission lines to changes in the continuum. One then uses these times to determine the distances of the emitting regions from the SMBH and make SMBH mass estimates using the virial theorem. Our current picture of the BLR involves a large number of clouds with a range of distances from the SMBH arranged in some type of flattened or toroidal distribution. The emission we see from the AGN depends on our viewing angle with respect to the orientation of the BLR.

Unfortunately, reverberation mapping has only been done for a small number of AGN, and doing it for new objects requires observations over a period of many years. In order to study AGN and obtain SMBH mass estimates without reverberation mapping data, we need methods that work with single-epoch observations. Many studies aimed at this have used the photoionization code \Cloudy \citep{ferland13} to compare observations to BLR models \citep[e.g.,][]{hamann93,baldwin95,wang12}. Emission lines used in these studies include FUV and NUV lines such as \Osix $\lambda\lambda$1032, 1038, \Nfive $\lambda\lambda$1239, 1243, Ly$\alpha$, \ion{C}{4} $\lambda\lambda$1548, 1551, and \ion{Mg}{2} $\lambda\lambda$2796, 2803. 

The high resolution provided by the Cosmic Origins Spectrograph (COS) on the \textit{Hubble Space Telescope} allows similar models to be applied to EUV emission lines \citep{green12}. In this paper, we use \Cloudy to model COS spectra of four individual AGN, along with a composite UV and EUV spectrum of 159 AGN from \citet{stevans14}. We use our models to constrain the properties of the BLR, and we compare the incident UV continuum favored by our models to that measured from our spectra. We describe our object selection and emission-line measurements in Section \ref{data}. Details of our \Cloudy photoionization grids are given in Section \ref{cloudy}. In Section \ref{models} we describe our two BLR models and fit them to our measured fluxes. We discuss the implications of our results and future work in Section \ref{discuss} and summarize our conclusions in Section \ref{sum}. Throughout this paper, spectral indices $\alpha_\nu$ (also called $\alpha_{UV}$) will be given in terms of frequency, with the relationship to flux $F_\nu \propto \nu^{\alpha_\nu}$. Spectral indices in terms of wavelength ($\alpha_\lambda$) can be determined using the relationship $\alpha_\nu + \alpha_\lambda = -2$. All atomic data in this paper come from the NIST Atomic Spectra Database\footnote{http://www.nist.gov/pml/data/asd.cfm}.

\section{EMISSION LINE MEASUREMENTS}\label{data}

\subsection{Object Selection and Data Reduction}

AGN were selected from among those with publicly available COS G130M (1133-1468 \AA) and G160M (1383-1796 \AA) spectra covering the restframe wavelength range 760-1100 \AA. This corresponds to a redshift range of $0.49 \lesssim z \lesssim 0.63$. This range contains strong emission lines from ions with ionization energies ranging from 13.6 eV for \Hone or \Oone to 239.1 eV for \Neeight, and excitation energies up to 16.2 eV for \Nfour $\lambda765$. In addition it contains relatively line-free windows at $\sim 850$ \AA{} and 1100 \AA{} for continuum fitting. As the goal of this study was to determine the feasability of using \Cloudy models to constrain BLR properties and intrinsic AGN spectral indices, no attempt was made to compile a complete sample of AGN within this redshift range. Instead, a small sample of radio-quiet AGN was chosen with median S/N per pixel of at least three. In addition, only spectra without Lyman limit or partial Lyman limit systems were selected. These requirements provided a sample with a broad range of observed UV spectral indices ($\alpha_{UV}$) and spectra allowing for accurate line flux measurements. Four AGN were chosen that met this criteria: HE~0226-4110, HS~1102+3441, SDSS~J124154.02+572107.3, and HE~0238-1904. Properties of the AGN and their spectra are given in Table \ref{tb1}.

Two additional AGN were chosen with spectra covering the restframe wavelength range 950-1300 \AA. This range contains emission from both \Nthree $\lambda991$ and \Nfive $\lambda1240$. Having lines from different ions of the same element makes it easier to test for non-solar metallicity. Chemical enrichment models of AGN also identify nitrogen as the elemental abundance most sensitive to metallicity variations \citep{hamann93}. As before, the spectra were selected to have high median S/N and relatively clean emission lines. The two AGN chosen were RXJ~2154.1-4414 and SDSS~J132222.68+464535.2. Their properties are given in Table \ref{tb1}. Throughout the paper, these two AGN will be referred to as the ``\Nfive spectra'' to distinguish them from the ``main spectra'' described in the previous paragraph.

Single epoch COS G130M and G160M spectra were obtained from the Mikulski Archive for Space Telescopes (MAST)\footnote{http://archive.stsci.edu/}. The properties of COS are described in \citet{green12} and \citet{osterman11}. Individual spectra were coadded using the IDL routines available on the COS Tools website\footnote{http://casa.colorado.edu/\textasciitilde danforth/science/cos/costools.html} and described in \citet{coadd} and \citet{coadd2}. The spectra were then corrected for Galactic extinction using a \citet{fitzpatrick99} reddening law with dust maps from \citet{dustmap} and $R_V = 3.1$. The spectra were shifted to the rest frame using AGN redshift data from NED\footnote{http://ned.ipac.caltech.edu}. The redshift determined from AGN emission lines is dependent on the lines used. Individual lines can have relative velocity offsets of several thousand km s$^{-1}$, with higher ionization lines being blueshifted relative to lower ionization ones \citep{gaskell82,corbin90,gaskell13}. Our analysis is not dependent on knowing exact AGN redshifts, but these velocity shifts will be important in our discussion of the \Nthree 991 \AA{} line. Finally, all strong absorption lines were masked from the spectra, along with geocoronal emission from Ly$\alpha$ and \Oone{} $\lambda$1304.

\begin{figure*}
\plotone{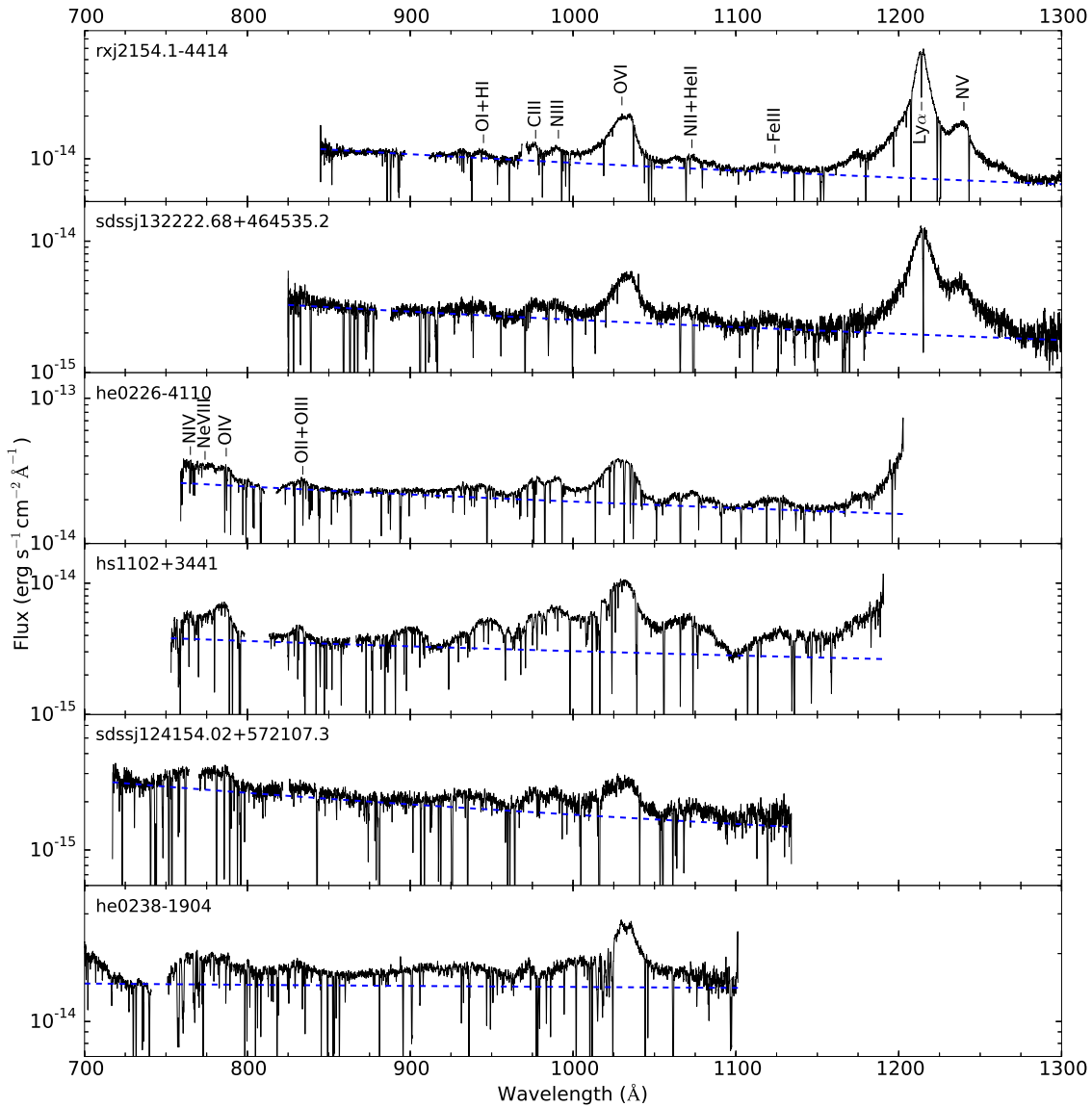}
\caption{Rest-frame COS G130M/G160M spectra of the four AGN in our sample along with their continuum fits from \citet{stevans14}. Geocoronal emission lines have been masked from the spectra, but absorption lines have not.\label{spectra}}
\end{figure*}

In order to fit a continuum to each spectrum, we needed to identify regions of the spectrum that are relatively free of contamination from emission lines. \citet{composite1} list line-free continuum windows over a wide range of UV wavelengths. The relevant windows for the main spectra are 715-730 \AA{}, 860-880 \AA{}, and 1090-1105 \AA{}. Continuum windows were chosen as widely seperated in wavelength as possible, to provide long baselines for the fits, while avoiding the ends of the spectra where the S/N is poor. All four main spectra used the window from 1090-1105 \AA{} for one side of the continuum fit. HE~0226-4110 and HS~1102+3441 used the 860-880 \AA{} window for the other side of the fit, while the other two spectra used the 715-730 \AA{} window. The \Nfive spectra have additional continuum windows at 1280-1290 \AA{} and 1315-1325 \AA{} \citep{composite1}. The fit to RXJ~2154.1-4414 used the windows at 860-880 \AA{} and 1315-1325 \AA{}, while the fit to SDSS~J132222.68+464535.2 used 860-880 \AA{} and 1090-1105 \AA. The MPFIT IDL routine \citep{mpfit} was used to fit a power-law continuum to the chosen windows. The spectral indices for the fits are given in Table \ref{tb1} and range from $\alpha_{UV} = -1.90$ to $-0.60$. Spectra for the six AGN along with their continuum fits are shown in Figure \ref{spectra}.

\begin{deluxetable*}{lllrr}
\tablewidth{0pt}
\tablecaption{AGN Properties\label{tb1}}
\tablehead{\colhead{Name} & \colhead{Type\tablenotemark{a}} & \colhead{$z_{\text{AGN}}$\tablenotemark{a}} & \colhead{$\alpha_{UV}$\tablenotemark{b}} & \colhead{S/N\tablenotemark{c}}}
\startdata
RXJ 2154.1-4414 & QSO & 0.344 & $-0.676\pm0.001$ & 9.2 \\
SDSS J132222.68+464535.2 & QSO & 0.374873 & $-0.646\pm0.002$ & 3.2 \\
HE 0226-4110 & QSO & 0.493368 & $-0.930\pm0.003$ & 13.4 \\
HS 1102+3441 & QSO & 0.508346 & $-1.203\pm0.008$ & 6.6 \\
SDSS J124154.02+572107.3 & QSO & 0.583471 & $-0.598\pm0.003$ & 3.8 \\
HE 0238-1904 & QSO & 0.631 & $-1.902\pm0.005$ & 9.7 \\
\enddata
\tablenotetext{a}{AGN type and redshift data from NED}
\tablenotetext{b}{Spectral indices with respect to frequency ($F_\nu \propto \nu^{\alpha_\nu}$)}
\tablenotetext{c}{Median S/N per pixel across the coadded spectrum}
\end{deluxetable*}

\newpage
\subsection{Emission Line Flux Measurements}\label{flux_meas}
The spectral range for the main spectra contains several strong emission lines, the most prominent being the \Osix $\lambda\lambda$1032, 1038 doublet which is typically blended with Ly$\beta$ $\lambda$1026. Blueward of the \Osix{} doublet are two weaker emission lines: \Nthree $\lambda$991 (which includes a contribution from \Hetwo $\lambda$992) and \Cthree $\lambda$977. At the blue end of our spectral range is a set of several emission lines: \Nfour $\lambda$765, \Neeight $\lambda\lambda$770, 780, and a triplet of \Ofour lines with mean wavelength 789 \AA{}. These lines are too heavily blended to easily separate, so they are fit as a single entity. We will refer to this blend as the \Neeight complex, as the emission from the \Neeight doublet is the dominant component. The final emission line that we measure is the blend of \Otwo $\lambda$833 and \Othree $\lambda$834.

There were two regions of emission in our spectra that were excluded from the final flux measurements. The first of these is the region redward of the \Osix doublet. There are emission lines from \Ntwo and \Hetwo predicted at 1085 \AA{}. However, as can be seen in Figure \ref{spectra}, the actual peak in the emission occurs closer to 1070 \AA{}. Our \Cloudy models do not predict any significant emission lines near this wavelength, meaning that either the \Ntwo and \Hetwo emission is blueshifted strongly relative to the other emission lines, or the emission is from an unknown source. A strong blueshift of the low ionization \Ntwo and \Hetwo relative to the higher ionization \Osix would be inconsistent with the current understanding of AGN emission line shifts. The second region spans the wavelength range from $\sim870$ to 950 \AA{}. This region is produced by the Lyman series and a large number of \Oone lines, along with \Ctwo $\lambda904$. The data processing and analysis difficulties of including this many lines in our \Cloudy models led us to exclude this region.

The \Nfive spectra overlap in wavelength with the main sample, containing the \Osix $\lambda\lambda$1032, 1038, \Nthree $\lambda$991, and \Cthree $\lambda$977 lines. The emission in the red half of the spectra is dominated by Ly$\alpha$ $\lambda$1216. Several emission lines exist on the wings of Ly$\alpha$. They include the \Nfive $\lambda\lambda$1239, 1243 doublet, which we measure. \Cthree $\lambda$1175 and \Sitwo $\lambda$1263 are also present, but we omit them from this study as they are weaker and the errors associated with deblending them from the Ly$\alpha$ wings would be high.

\begin{figure*}
\plotone{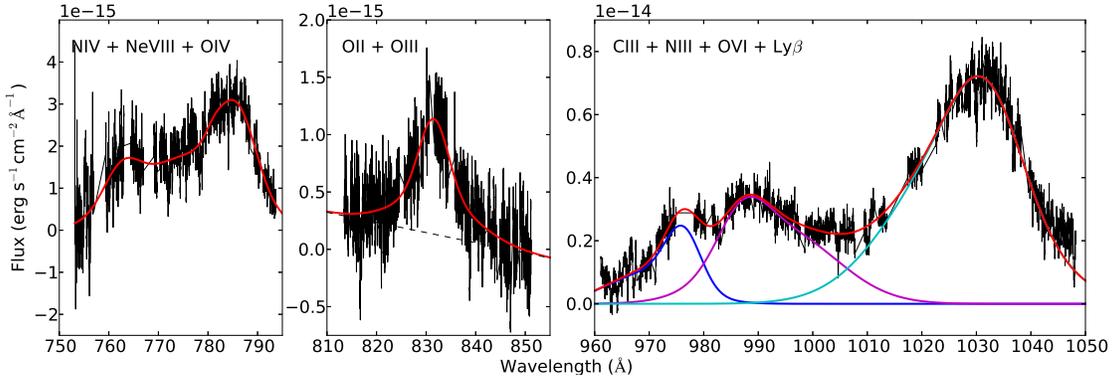}
\caption{Fits to all of the emission lines measured in HS 1102+3441 in the accessible band. The fit to the \Otwo + \Othree blend at 833 - 834 \AA{} includes a linear pseudocontinuum.\label{fig2}}
\end{figure*}

After subtracting the continuum fits from our spectra, the emission lines were fit with sums of Gaussians using MPFIT. The \Osix, \Nthree and \Cthree lines were fit simultaneously, while the \Otwo + \Othree blend and \Neeight complex were each fit independently. No significance should be given to the individual Gaussian components; they are only a means of approximating the line shapes. In order to restrict the fits as little as possible, the only constraints placed on the Gaussians were that their amplitudes be positive, and that their peaks fall within a fixed distance from restframe wavelength of the line, typically 5 \AA{}. In particular, the wavelength offsets and widths of the different lines were not fixed relative to each other, allowing for measurement of emission originating from different portions of the BLR. Additional Gaussian components were added to the line fits until the $\chi^2$ value of the fit ceased to improve significantly. The number of Gaussian components used ranged from one to four depending on the line and spectrum. For several lines, residual continuum remained after the continuum fits had been subtracted. In these cases, a constant or linear pseudocontinuum was fit along with the Gaussians. Line fluxes were calculated by summing the areas of the Gaussian components and are given in Table \ref{tab2}. Fits to all of the lines measured in HS~1102+3441 are shown in Figure \ref{fig2}. In two cases, line fluxes were unable to be determined. The bluest portion of the HE~0226-4110 spectrum was unusable, cutting off a portion of the \Neeight complex. A lower bound on the \Neeight complex flux was calculated by directly integrating over the clean portion of the complex, but this region was too small to provide a useful constraint on the \Cloudy models. Due to low S/N, a Gaussian failed to fit the \Otwo + \Othree blend in SDSS~J124154.02+572107.3 any better than a linear pseudocontinuum alone.

\subsection{Flux Error Sources}

Statistical errors in the line fluxes for a fixed continuum fit were calculated by simple propagation of errors from the Gaussian fits. Typical values for these errors are a few percent of the total line flux, with a maximum value of $13.5\%$ for the \Nthree $\lambda$991 line in HE~0238-1904. However, the statistical errors are only one of several important sources of uncertainty. Additional sources of error include deblending of nearby emission lines, errors in the extinction corrections, and uncertainties in the continuum subtraction. As can be seen in Figure \ref{fig2}, the \Osix, \Nthree, and \Cthree lines are moderately blended. In the \Nfive spectra there is additional blending between Ly$\alpha$ and \Nfive. It is possible that the fits to these lines include emission that was actually caused by a different line. A rough estimate of this uncertainty was obtained by varying the number of Gaussian components and initial conditions of the fits to determine the variation in measured fluxes. We adopted the typical flux variations from this testing of $7.5\%$ for \Osix, $15\%$ for \Nthree and \Cthree, $2\%$ for Ly$\alpha$, and $20\%$ for \Nfive as our deblending errors. In all cases, this is the primary source of error for these lines.

Errors in the extinction correction affect both the continuum fits and the line fluxes. These errors have two sources: uncertainty in the strength of the extinction as measured by E(B-V) and uncertainty in the form of the extinction law. We measured these errors separately, varying the E(B-V) values by their 1$\sigma$ uncertainties of $\pm16\%$ \citep{schlegel98} and the $R_V$ values to 2.8 and 4.0 from the standard 3.1. To calculate the change in line fluxes, the spectra were corrected for extinction with both the standard and modified parameters. Continua were then fit and subtracted, and the difference of the two spectra was calculated. This difference was integrated over the wavelength range of the line to find the change in flux. For lines which were fit simultaneously or along with a pseudocontinuum, the value of the difference in each pixel was multiplied by the ratio of the fit to the line of interest to the total fit for that pixel. Finally, the integrated difference was multiplied by the ratio of the total line flux to the line flux from unmasked pixels to compensate for the masking. The resulting flux uncertainties are nearly symmetric in all cases and so were treated as symmetric. The values are reported in Table \ref{tab2}. The errors due to uncertainty in E(B-V) are typically at the percent level and are comparable to the statistical errors in the fits. The errors due to uncertainty in $R_V$ are lower, often by as much as an order of magnitude, but they grow in importance toward the blue end of the spectral range.

The uncertainties in extinction parameters that we use for our estimates are typical for the Galaxy as a whole. However, \citet{peek13} find that at the high Galactic latitudes where our sources lie, extinction corrections are even less certain. They find that no single value of $R_V$ is valid over the entire high-latitude sky, and that accurate $R_V$ values may differ substantially from the commonly adopted value of $R_V = 3.1$. In addition, our analysis only accounts for Galactic extinction. There is almost certainly additional extinction due to the local environment of the AGN, as well as their host galaxies, which we do not correct for. \citet{gaskell04} found that typical intrinsic AGN extinction curves are flatter than Galactic extinction in the UV. Their proposed reddening curve does not extend shortward of Ly$\alpha$, so we cannot use it to correct for intrinsic extinction in our spectra. As a result of these issues, the extinction errors we calculate most likely underestimate the true values. Uncertainties in extinction can have an important effect on the continuum fits and line flux measurements. However, when calculating equivalent widths these effects offset each other, reducing their importance. They are unlikely to affect the total errors in the equivalent width for the lines with blending uncertainties, but for \Otwo + \Othree and the \Neeight complex they are potentially important.

Due to the relatively high S/N of our spectra, statistical uncertainties in our continuum fits are small compared to other sources of error. The errors due to the extinction corrections are handled implicitly by the previously described calculations, as the continuum is refit and subtracted for each modified extinction value before the spectra are differenced. The biggest remaining source of error in our continuum fits is due to line contamination of our continuum windows. Although these windows were chosen to be far from any strong emission lines, the extremely broad nature of BLR emission lines and the large number of lines involved means that some residual contamination is unavoidable. Unfortunately, the nature of this contamination makes it extremely difficult to quantify. This likely represents the most important unquantified source of error in our flux values.

All measured error sources are summarized in Table \ref{tab2}. Individual errors were added in quadrature to obtain the total error values. These values range from $\pm2.1\%$ for the \Neeight complex in SDSS~J124154.02+572107.3 to $\pm21.5\%$ for the \Nfive $\lambda$1240 line in SDSS~J132222.68+464535.2, with a mean of $\pm11.2\%$.

\subsection{Composite Spectrum}

Along with our individual spectra, we also measured line fluxes for the composite spectrum of 159 AGN spectra created by \citet{stevans14}. The procedure for creating the composite spectrum is decribed in \citet{composite1}. The composite covers a wavelength range from $\sim450$ \AA{} to $1800$ \AA{}, and its continuum is fit with a broken power law with EUV continuum spectral index of $\alpha_\nu =-1.41\pm0.15$ and an FUV index of $\alpha_\nu=-0.83\pm0.09$ \citep{stevans14}. The break between the two power law indices occurs at $\sim 1000\pm25$ \AA. The physical explanation for this continuum fit is likely a gradual change in the spectrum rather than a true break. We measured fluxes for a total of 20 different lines and blends over the range $600-1800$ \AA{}. The portion of the composite shortward of $600$ \AA{} had S/N too low for accurate flux measurements. Table \ref{tab3} gives the flux values for each of the lines along with the statistical errors in the fits to each line. Errors due to extinction corrections were not calculated, as our method would require that each spectrum in the composite be individually fit, which is beyond the scope of this paper.

\begin{deluxetable}{llrrr}
\tablewidth{0pt}
\tablecaption{Composite Spectrum Emission Line Fluxes\label{tab3}}
\tablehead{\colhead{Ion} & \colhead{Wavelength (\AA{})} & \colhead{Flux\tablenotemark{a}} & \colhead{Error\tablenotemark{a}} & \colhead{EW (\AA{})}}
\startdata
\ion{O}{3}] & 1663 & 1.84 & $\pm0.38$ & 3.38 \\
\Hetwo & 1640 & 12.91 & $\pm0.63$ & 23.36 \\
\ion{C}{4} & 1550 & 55.66 & $\pm0.68$ & 94.29 \\
\ion{N}{4}] & 1486 & 0.86 & $\pm0.17$ & 1.39 \\
\ion{Si}{4} + \Ofour & 1400 & 11.49 & $\pm0.33$ & 17.28 \\
\ion{C}{2} & 1335 & 1.03 & $\pm0.14$ & 1.46 \\
\ion{O}{1} + \ion{Si}{2} & 1305 & 3.23 & $\pm0.25$ & 4.48 \\
\ion{Si}{2} & 1263 & 0.40 & $\pm0.17$ & 0.53 \\
\Nfive & 1240 & 4.09 & $\pm0.36$ & 5.33 \\
Ly$\alpha$ & 1216 & 100.00 & $\pm0.82$ & 127.48 \\
\Cthree & 1175 & 0.48 & $\pm0.15$ & 0.59 \\
\Osix + Ly$\beta$ & 1033 & 25.34 & $\pm0.47$ & 26.69 \\
\Nthree & \phn991 & 1.99 & $\pm0.25$ & 2.01 \\
\Cthree & \phn977 & 3.38 & $\pm0.27$ & 3.40 \\
\Ctwo & \phn904 & 0.83 & $\pm0.39$ & 0.80 \\
\Otwo + \Othree & \phn833 & 3.45 & $\pm1.20$ & 3.14 \\
\Neeight Complex & \phn773 & 13.43 & $\pm1.39$ & 11.73 \\
\Othree & \phn702 & 2.58 & $\pm1.26$ & 2.13 \\
\Nthree & \phn686 & 5.13 & $\pm0.91$ & 4.16 \\
\ion{O}{5} & \phn629 & 8.49 & $\pm2.68$ & 6.55 \\
\enddata
\tablenotetext{a}{Values normalized to a Ly$\alpha$ flux of 100}
\end{deluxetable} 

\subsection{Variations Between Spectra}\label{var}

\begin{figure*}
\plotone{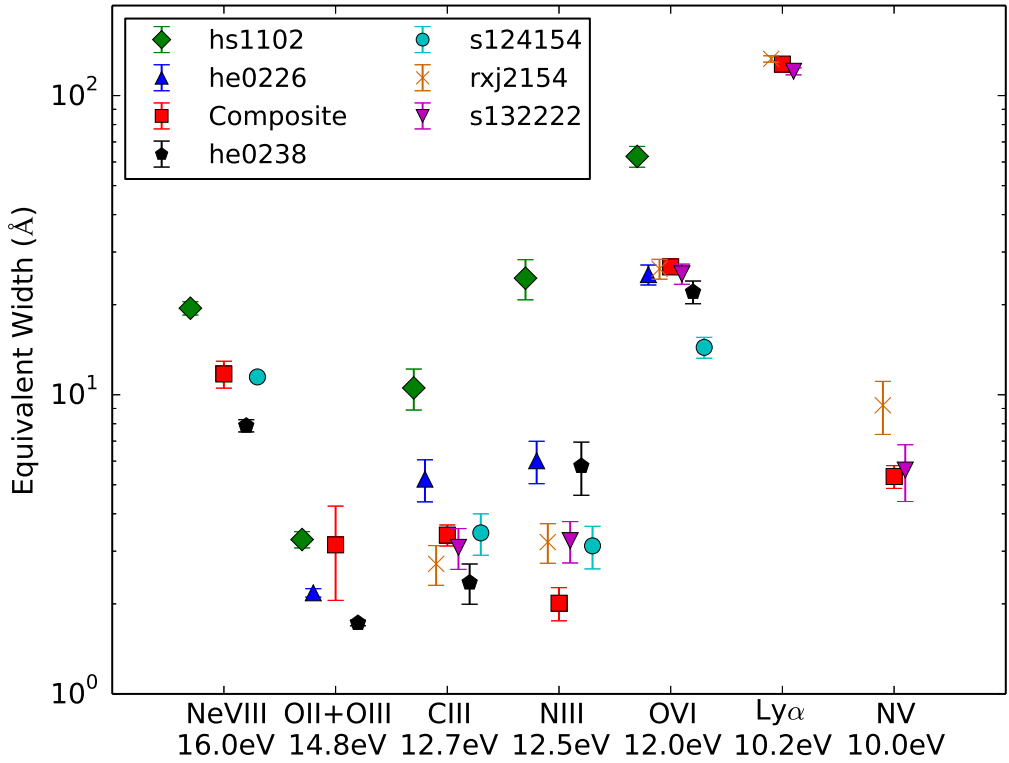}
\caption{Equivalent widths of all emission lines measured in our study. Excitation energy of the lines increases from right to left. Error bars indicate the total 1$\sigma$ error estimates for the fluxes. Horizontal offsets between AGN are added for clarity. Note in particular the large variation in \Nthree $\lambda$991 equivalent width values.\label{fig3}}
\end{figure*}

Due to differences in both intrinsic luminosity and distance, the best way to compare emission lines between spectra is in terms of equivalent widths ($W_\lambda$), calculated as 
\begin{equation}
W_\lambda = \frac{F_{\text{line}}}{F_{\text{cont}}}\; ,
\end{equation}
where $F_{\text{line}}$ is the total line flux (erg s$^{-1}$ cm$^{-2}$) and $F_{\text{cont}}$ is the flux in the continuum fit at the wavelength of the line (erg s$^{-1}$ cm$^{-2}$ \AA$^{-1}$). For blends, the wavelength of the dominant component is used. Equivalent widths for our measured lines are plotted in Figure \ref{fig3}, including values for the composite spectrum. Overall, the equivalent widths for a given emission line vary by a factor of a few between spectra, except for \Nthree $\lambda$991 which varies by over a factor of 10. The values from the composite spectrum fall within the range of the individual spectra, again with the exception of \Nthree. Equivalent widths from different spectra are typically not consistent within their 1$\sigma$ errorbars. This means that attempting to increase our available sample by stacking low S/N spectra from several AGN would not give useful results, although the composite spectrum might still provide information on average AGN properties. The large variation in \Nthree equivalent widths will be examined in more detail in Section \ref{Nthree}.

\section{CLOUDY MODEL GRIDS}\label{cloudy}

All photoionization models were calculated using version 13.00 of \Cloudy, which is described in \citet{ferland13}. Inputs to the models include the geometry, physical and chemical properties of the emitting ``clouds'' of gas, along with the properties of the incident radiation. The output parameters relevant to this paper are the intensities of the various emission lines produced in the clouds. Individual clouds have constant density and a fixed total hydrogen column density of $10^{23}$ cm$^{-2}$. The choice of constant density means that there will be pressure differentials across individual clouds. Constant pressure clouds have lower densities in the more highly ionized regions, affecting the line ratios between low and high ions for a given cloud. The cloud's column density affects the size of the range in cloud density and ionizing flux for which a given line can be efficiently produced, with higher column densities providing larger ranges \citep{korista97_1}. In our analysis, these effects will be more important for the single-component models than for the locally optimally emitting cloud models, where we average over the properties of a broad range of clouds (Section \ref{models}). The clouds were assumed to be small enough relative to the scale of the BLR that they could be treated as simple 1D slabs irradiated on one side, with a normally incident flux. The models assumed that there was no shielding of the continuum source by other clouds. The primary effect of shielding by other clouds is a change in the relationship between ionizing flux and distance from the continuum source \citep{gaskell07}. The motions within the cloud were assumed to be entirely thermal, with no turbulent or wind contributions to the velocities. Finally, the clouds were assumed to have solar metallicity and abundance ratios, although as will be discussed in Section \ref{Nthree} a higher metallicity may be required to accurately model the \Nthree $\lambda$991 emission.

Models were computed for a grid of cloud hydrogen number densities ($n_H$) and incident fluxes of hydrogen-ionizing photons ($\Phi_H$). These grids represent clouds with a range of densities and distances from the continuum source, with $\Phi_H \propto r^{-2}$ functioning as a proxy for distance. The model grids cover a density range of $7 \le \log n_H (\text{cm}^{-3}) \le 14$ with 0.25 dex spacing. The upper limit is set primarily by the limitations of \Cloudy, as some of the atomic calculations become uncertain at $n_H\approx 10^{15} \text{ cm}^{-3}$ \citep{ferland13}. The lower density limit is determined by the absence of forbidden lines in the BLR \citep{baldwin95}. The range of ionizing flux in the grid is $17 \le \log \Phi_H (\text{cm}^{-2}\text{ s}^{-1}) \le 24$ with 0.25 dex spacing, which was chosen to cover the regions of peak emission for our BLR emission lines \citep{korista97_1}.

The covering factor of BLR clouds is an important parameter in our \Cloudy models. The lack of broad absorption lines in many AGN spectra rules out complete covering of the continuum source by BLR clouds. The primary effect of covering factor is a linear rescaling of the emission line equivalent widths from the 100\% covering case. However, comparing \Cloudy models with different covering factors indicates that secondary effects can introduce scatter of 5-10\% around this simple rescaling for a given line and grid point. As a result, a specific value needs to be chosen for our model grids. We adopt 40\% covering as a value consistent with the literature \citep{dunn07,gaskell07}. The success of our models in fitting many of our AGN spectra (Section \ref{models}) indicates that this is a reasonable choice. However, the high equivalent widths of the lines in HS~1102+3441 likely require a larger covering factor.

\subsection{Input Continuum Shapes}

\begin{figure*}
\plotone{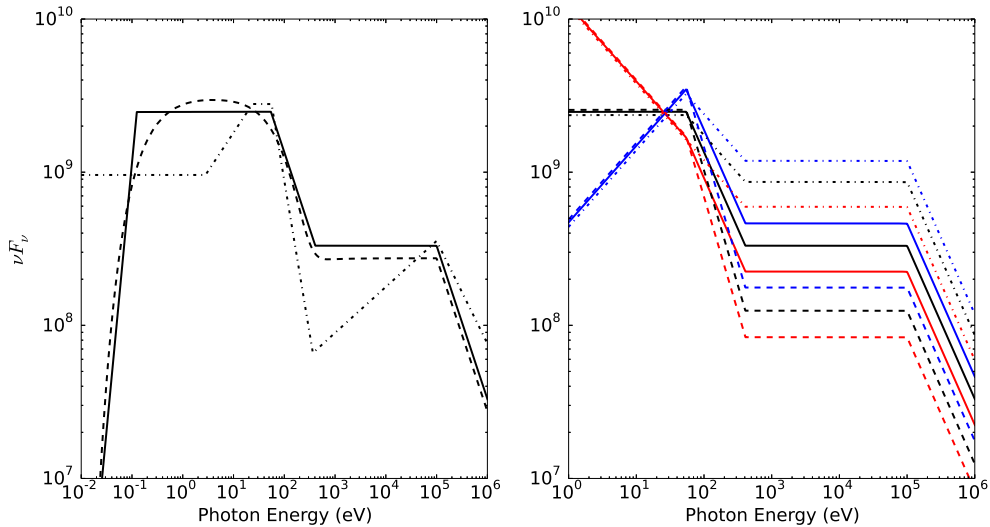}
\caption{Plots of incident continuum SEDs for the different \Cloudy grids. The left panel compares $\nu F_\nu$ for the grid with $\alpha_{UV}=-1.0$ and $\alpha_{\text{trans}}=-2.0$ (solid) to the template AGN spectrum from \citet{mathews87} (dash-dotted) and the input continuum from \citet{korista97_1} with bump temperature $10^6$ K and $\alpha_{ox} = -1.4$ (dashed), where $\alpha_{ox}$ is the spectral index from 2500 \AA{} to 2 keV. The right panel compares all nine of our incident SEDs at UV and x-ray energies. Colors denote different UV spectral indices, with $\alpha_{UV} = -1.5$, $-1.0$, and $-0.5$ as red, black, and blue, respectively. Line styles denote different 4 - 30 Ryd spectral indices, with $\alpha_{\text{trans}}=-2.5$, $-2.0$, and $-1.5$ as dashed, solid, and dash-dotted, respectively.\label{inc}}
\end{figure*}

The shape of the spectral energy distribution (SED) of the incoming radiation plays a central role in determining both the ionization states and electron excitations in our models. In order to constrain the UV spectral indices of our AGN sample we must also be able to easily vary the relevant regions of the SED. We chose an input SED that consists of a series of broken power laws. Two portions of our SED have spectral indices that vary between the different \Cloudy grids. The IR to UV continuum from $9.12\e{-3}$ Ryd (10 $\mu$m) to 4 Ryd has a frequency spectral index $\alpha_{UV}$ with a default value of $-1.0$. From 4 to 30 Ryd there is a UV to X-ray transition region with index $\alpha_{\text{trans}}$ and a default value of $-2.0$. The X-ray portion of the spectrum has an index of $\alpha_{x} = -1.0$ and extends to $7350 \,\text{Ryd} \approx 100$ keV. At higher energies the spectrum falls off with a power law index of $-2.0$, while longward of 10 $\mu$m the SED also decreases rapidly with an index of $2.5$. Our default values for $\alpha_{UV}$ and $\alpha_{\text{trans}}$ produce a spectrum that closely matches the baseline spectrum in \citet{korista97_1} for wavelengths shorter than 1 $\mu$m. The two spectra are shown side by side in the left panel of Figure \ref{inc}.

In order to investigate the effects of SED variations, we computed several \Cloudy grids with varying values of $\alpha_{UV}$ and $\alpha_{\text{trans}}$. Values of $\alpha_{UV}$ in the literature cover a wide range. The template AGN SED in \citet{mathews87} has a spectral index of $-0.5$ between 2.8 and 23.7 eV. The AGN composite spectrum from \citet{scott04} found $\alpha_{UV} = -0.56$, but is likely much harder than the true value due to broad EUV emission lines being treated as continuum. Other AGN composite spectra have softer UV continua, with \citet{telfer02} finding $\alpha_{UV} = -1.57\pm0.17$ for radio quiet AGN, and \citet{composite1} finding $\alpha_{UV} = -1.41\pm0.20$ for 22 AGN, which was updated to $\alpha_{UV} = -1.41\pm0.15$ in \citet{stevans14} for the expanded composite of 159 AGN spectra. Individual AGN show an even wider range of $\alpha_{UV}$. We used $\alpha_{UV}$ values of $-0.5$, $-1.0$, and $-1.5$, and $\alpha_{\text{trans}}$ values of $-1.5$, $-2.0$, and $-2.5$ for a total of nine \Cloudy grids. The right panel of Figure \ref{inc} compares these nine input spectra for UV and X-ray energies. This range of spectral indices produces significant variation over the ionization energies relevant to the measured lines in our spectra. The lowest energy is 13.6 eV for ionizing \Hone and producing \Otwo, while the highest is 239 eV for ionizing \Neeight to \ion{Ne}{9}. At fixed $\Phi_H$, the ratio of the number of photons ($F_\nu/h\nu$) at 239 eV/13.6 eV is 17.6 times higher for the hardest spectrum ($\alpha_{UV} = -0.5$, $\alpha_{\text{trans}} = -1.5$) than for the softest ($\alpha_{UV} = -1.5$, $\alpha_{\text{trans}} = -2.5$).

\begin{figure*}
\plotone{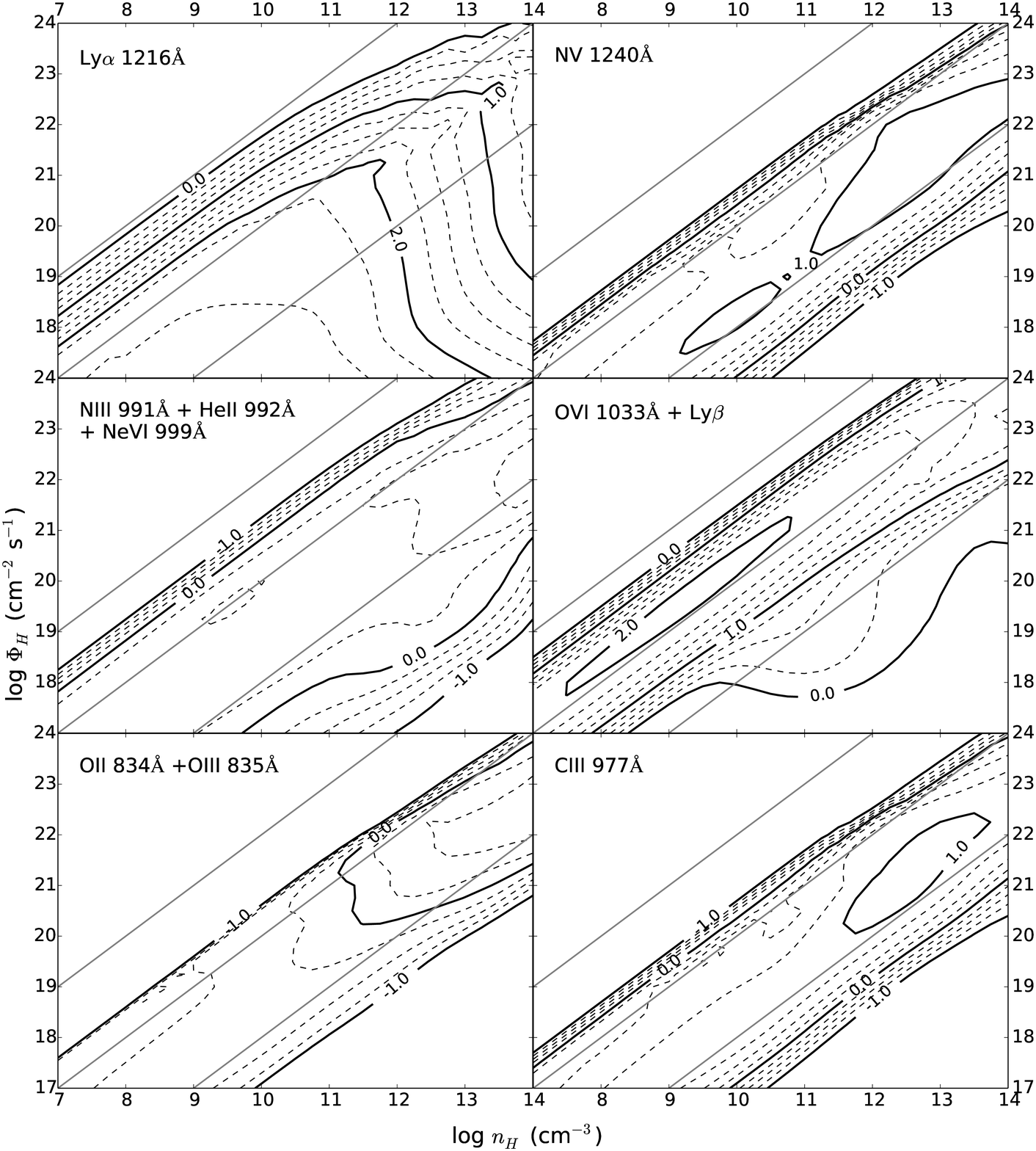}
\caption{Logarithmic contour plots of the equivalent widths of all measured lines except for the \Neeight complex. The incident spectrum for this grid had $\alpha_{UV} = -1.0$ and $\alpha_{\text{trans}} = -2.0$. The gray diagonal lines are photoionization parameter contours. Note that all of the equivalent widths except for Ly$\alpha$ peak along lines of constant photoionization parameter.\label{fig5}}
\end{figure*}

\begin{figure*}
\plotone{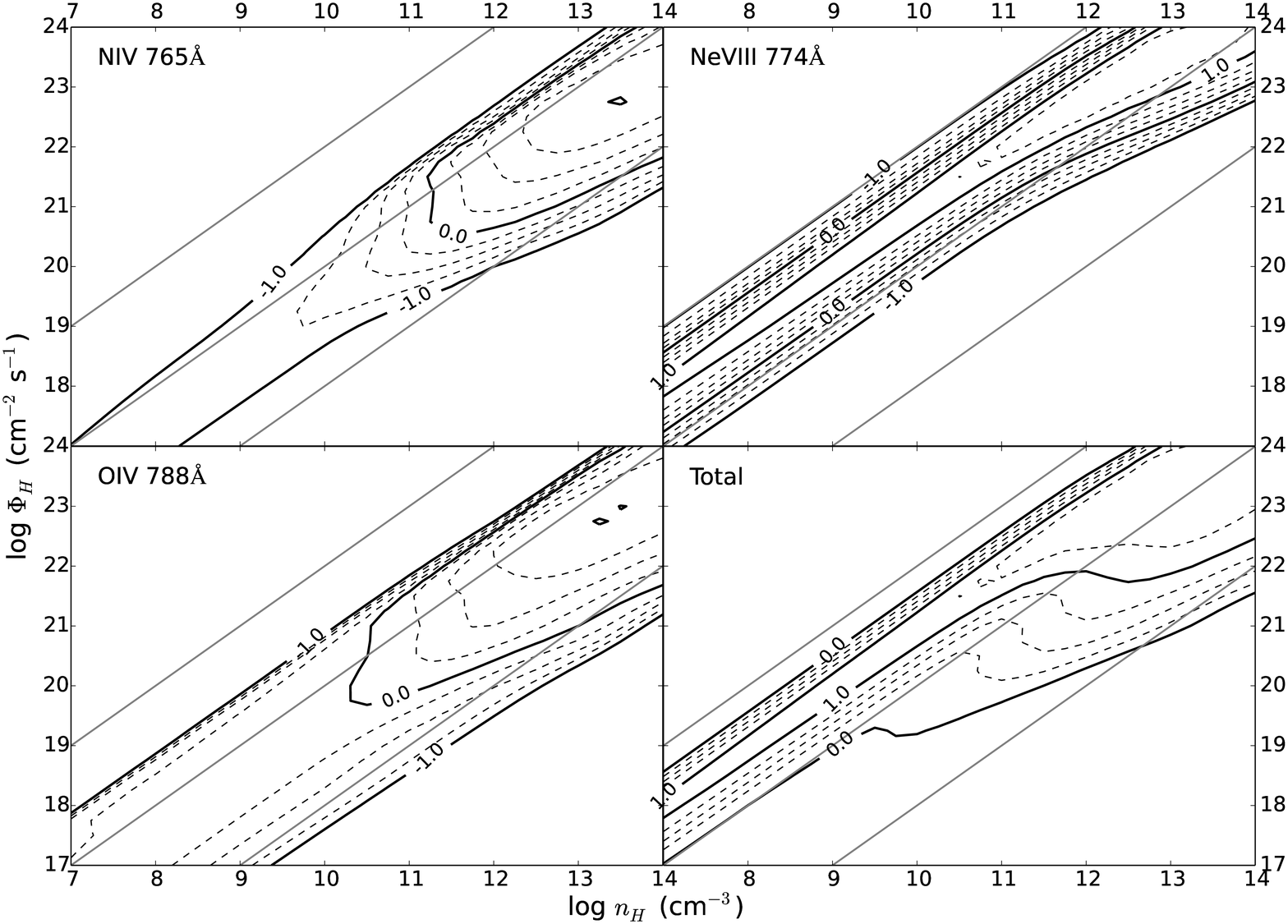}
\caption{Logarithmic contour plots of the equivalent widths of the three lines contributing to the \Neeight complex, along with the total for the complex as a whole. The incident spectrum for this grid had $\alpha_{UV} = -1.0$ and $\alpha_{\text{trans}} = -2.0$. The gray diagonal lines are photoionization parameter contours. The strongest component of the complex is the \Neeight doublet at $770$ \AA{} and $780$ \AA{} (weighted average $774$ \AA{}). As in Figure \ref{fig5}, the equivalent widths peak along lines of constant photoionization parameter.\label{fig6}}
\end{figure*}

\subsection{Emission Line Equivalent Widths}\label{EW}

The primary output of our \Cloudy models is a set of fluxes for our emission lines at each grid point. For blends of lines, we sum the \Cloudy model fluxes for each individual line. The blends consist of the individual components listed in the first paragraph of Section \ref{flux_meas}. In addition, our \Cloudy models predicted emission from a set of five \ion{Ne}{6} lines in the wavelength range 993 - 1010 \AA{} with a combined maximum equivalent width of $\sim15\%$ that of \Nthree $\lambda$991. Figure \ref{fig2} shows that for HS~1102+3441, the majority of the flux in this wavelength range is attributed to \Nthree, and the same is true for the other spectra. We add the \ion{Ne}{6} flux predicted by \Cloudy to the \Nthree $\lambda$991 flux.

Due to the varying intensity of the incident continuum between AGN, the line flux values are most usefully interpreted in the form of equivalent widths ($W_\lambda$), calculated as described in Section \ref{var}. Figures \ref{fig5} and \ref{fig6} show contour plots of $W_\lambda$ for all of our measured emission lines, along with the individual components of the \Neeight complex. The strongest contribution to the \Neeight complex comes from the \Neeight doublet, with a maximum $W_\lambda$ $\approx2$ times that of the other two lines. For all lines except Ly$\alpha$, the equivalent width values peak along lines of constant photoionization parameter ($U \equiv \Phi_H/cn_H$) as predicted by simple photoionization calculations. For single lines, this peak occupies a narrow range of $U$, while for blends the peak is broader as the individual components peak at different values.

An important difference between the emission lines is how much their equivalent widths vary along lines of constant $U$. The \Nfive $\lambda$1240, \Osix $\lambda$1035, and \Cthree $\lambda$977 equivalent widths show little variation from their maximum values at constant $U$, while \Nfour $\lambda$765, \Ofour $\lambda$788, and the \Otwo + \Othree 834 \AA{} blend all vary by over 1 dex. All of the lines with significant variation favor clouds with high density ($\log n_H \geq 12.0$) and ionizing flux ($\log \Phi_H \geq 22.0$). This is likely because the lines have high excitation energies ($\sim15$ eV), and only these clouds have electron temperatures high enough to sufficiently excite them. The line flux of a collisionally excited line is proportional to the excitation rate coefficient ($C_{12}$) given by 
\begin{equation}
C_{12} \propto \Omega_{12}T_e^{-1/2}e^{-E_{12}/kT_e}\; , 
\end{equation}
where $\Omega_{12}$ is the collision strength, $T_e$ is the electron temperature, and $E_{12}$ is the excitation energy of the line. The exponential dependence of the line flux on $T_e$ means that only clouds with electron temperatures within a factor of a few of the excitation energy of a line will produce significant emission. At 16.2 eV, \Nfour $\lambda$765 has the highest excitation energy of the lines in our sample, which corresponds to an excitation temperature $T_\text{exc} = 1.88\e{5}$ K. These high excitation temperatures are one of the primary advantages of using EUV lines, as most of the lines that have been studied at longer wavelengths cannot provide similar temperature constraints.

It is impossible to define a single electron temperature for the emitting region of a specific line in our \Cloudy grids. The flux from each line comes from clouds with a range of $n_H$ and $\Phi_H$, and $T_e$ additionally varies within each individual cloud. It is possible, however, to calculate values of $T_e$ that may be typical for a given line. We identified the grid points in our model with $\alpha_{UV} = -1.0$ and $\alpha_{\text{trans}} = -2.0$ that produced the peak equivalent widths of the \Nfive $\lambda\lambda$1239, 1243 doublet, the \Otwo + \Othree 834 \AA{} blend, the \Neeight complex, and the \Osix $\lambda\lambda$1032, 1038 doublet. The peak emitting clouds for \Nfive, \Otwo + \Othree, and the \Neeight complex had high density and ionizing flux, with $(\log n_H, \log \Phi_H) = (12.25,\, 22.0)$, $(14.0,\, 22.5)$, and $(13.0,\, 23.25)$, respectively, while the \Osix emission had lower values, with $(\log n_H, \log \Phi_H) = (9.0,\, 19.5)$. For each of these grid points, we used \Cloudy to calculate the distributions of electron temperature and line emission with respect to depth into the cloud. We then calculated an emission-weighted mean $T_e$ for the cloud producing the peak emission of each line. The mean values of $T_e$ for the \Otwo + \Othree blend were 37,600~K for \Othree and 33,900~K for \Otwo, which are $22\%$ and $20\%$ of the respective excitation temperatures ($T_\text{exc}$) of the lines. The \Neeight complex emission originated from three distinct regions of the cloud due to the differing ionization energies of the ions involved. The \Nfour and \Ofour lines had mean $T_e$ of 50,400 and 54,700~K, respectively, which correspond to $27\%$ and $30\%$ of their excitation temperatures. The mean $T_e$ for \Neeight was significantly higher at 94,000~K, or $51\%$ of its $T_\text{exc}$. This difference means that the upper states of the \Neeight transition are more easily excited than the other lines in the complex, and is likely one of the main reasons why the \Neeight emission dominates the total flux. The \Nfive gas is cooler, with a mean $T_e$ of 36,000~K, or $31\%$ of $T_\text{exc}$. The mean $T_e$ for \Osix is lower still at 35,700~K, which corresponds to $26\%$ of $T_\text{exc}$. With the exception of \Neeight, the peak emission from all of these lines comes from regions with $T_e$ between $20\%$ and $31\%$ of $T_\text{exc}$, which for higher energy lines corresponds to clouds with higher $n_H$ and $\Phi_H$. This is the primary reason why studying EUV emission lines in AGN is important: their emission is particularly sensitive to clouds with high density and ionizing flux that are not as well probed by lower $T_\text{exc}$ UV lines.

\section{MODELING OF MEASURED FLUXES}\label{models}

\subsection{Single-Component Model}

The first model we used to fit our \Cloudy grids to the measured line fluxes was a simple single-component model. For each $(\log{n_H}, \log{\Phi_H})$ pair in a grid, we used MPFIT to fit a scale factor to the \Cloudy line fluxes that best reproduced the measured values. The grid point with the minimum $\chi^2$ value was taken as our best-fit single-component model. None of our individual spectra were well fit by single-component models of all of their measured lines. The best fit across the main spectra was the grid with $\alpha_{UV} = -0.5$ and $\alpha_{\text{trans}} = -1.5$ fit to HE~0226-4110. This model had a minimum $\chi^2 = 10.02$ for 2 degrees of freedom (dof) with $\log{n_H}=14.0$ and $\log{\Phi_H}=23.25$. A broad trend across all of the fits was that the strength of the \Nthree $\lambda$991 line was underpredicted by the \Cloudy models. This was particularly true for HS~1102+3441, with its high ratio of \Nthree $\lambda$991 to \Cthree $\lambda$977 emission. The best-fit \Cloudy model for this spectrum predicted a \Nthree flux of $17\%$ of the measured value. For the other three spectra, the model fluxes ranged from $43\%$ to $74\%$ of the measured fluxes. Single-component models were also fit to all four of the main spectra without using the \Nthree fluxes. Two models provided a good fits to the HE~0226-4110 data. The first model had $\alpha_{UV}=-1.0$ and $\alpha_{\text{trans}} = -2.5$, giving $\chi^2 = 1.20$ for 1 dof with $\log{n_H}=13.5$ and $\log{\Phi_H}=23.0$; while the second had $\alpha_{UV}=-0.5$ and $\alpha_{\text{trans}} = -1.5$, giving $\chi^2 = 2.06$ for 1 dof with $\log{n_H}=14.0$ and $\log{\Phi_H}=23.25$. Even without the \Nthree line, no single-component models provided a good fit to the other three spectra.

\begin{figure}
\plotone{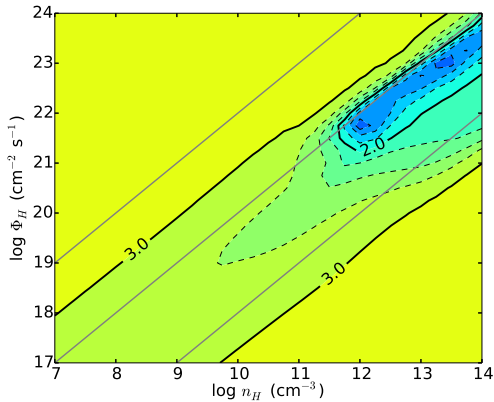}
\caption{Contour plot of $\log{\chi^2}$ for the \Cloudy model grid with spectral indices $\alpha_{UV} = -1.0$ and $\alpha_{\text{trans}} = -2.5$ from the fit to HE 0226-4110. The minimum $\chi^2$ values follow a line of constant $U$, with the lowest values occuring for clouds with the highest ionizing flux. Sample constant $U$ contours are shown by the gray diagonal lines.\label{single_comp}}
\end{figure}

One feature common across all of the single-component models for the main spectra was that the best-fit grid points had high values of $\log{n_H}$ and $\log{\Phi_H}$. Only two best-fit points had $\Phi_H < 10^{21}$ cm$^{-2}$ s$^{-1}$, and the majority had values $\ge 10^{22}$ cm$^{-2}$ s$^{-1}$. Figure \ref{single_comp} shows $\log{\chi^2}$ for the grid with $\alpha_{UV} = -1.0$ and $\alpha_{\text{trans}} = -2.5$ fit to HE~0226-4110 with the \Nthree flux excluded, but the general trends are typical for the fits as a whole. This plot shows trends similar to those of the equivalent width plots for individual lines discussed in Section \ref{EW}. For a fixed value of $U$, the measured emission lines are better fit by clouds with higher incident ionizing flux. This reinforces the importance of clouds with high electron temperatures for exciting EUV lines.

Fitting single-component models to the \Nfive spectra produced similar results. Neither spectrum had any good fits with all emission lines included, primarily because the \Nthree $\lambda$991 emission was underpredicted by $\sim 50\%$. There was no similar underprediction of the \Nfive $\lambda$ 1240 flux. With the \Nthree flux removed from the fits, SDSS~J132222.68+464535.2 had one good fit single-component model. This model had $\alpha_{UV}=-1.5$ and $\alpha_{\text{trans}} = -2.5$ with $\chi^2 = 3.33$ for 2 dof with $\log{n_H}=9.75$ and $\log{\Phi_H}=20.0$. No single-component models provided a good fit to RXJ~2154.1-4414 even with \Nthree excluded. The single-component models for the \Nfive spectra had significantly lower best-fit values of $\log{\Phi_H}$ than their counterparts in the main spectra. All best-fit grid points had $\log{\Phi_H} \le 10^{20.5}$ cm$^{-2}$ s$^{-1}$, with typical values closer to $10^{19.5}$ cm$^{-2}$ s$^{-1}$. This likely reflects the influence of Ly$\alpha$, which has its peak emission at much lower values of $\Phi_H$ than the other lines.

Single-component models were also fit to the composite spectrum. The models produced extremely poor fits to the full set of 20 measured emission lines,  with typical $\chi^2$ values of $>900$ for 18 dof. Models were also fit to the limited set of five emission lines measured in our main individual spectra. There were still no models that fit the composite spectrum well, but the quality of the fits was greatly improved. The best-fit model had spectral indices $\alpha_{UV} = -1.0$ and $\alpha_{\text{trans}} = -2.5$ with $\chi^2 = 17.67$ for 3 dof. Similarly to the individual spectra, the best-fit grid point had high ionizing flux, with $\log{n_H} = 12.5$ and $\log{\Phi_H} = 22.5$.

\subsection{Locally Optimally Emitting Cloud Model}

The second type of model we used was the locally optimally emitting cloud (LOC) model developed by \citet{baldwin95}. As discussed in Section \ref{EW}, for any given line significant emission occurs only over a narrow range of photoionization parameter (the ``optimally emitting'' clouds). By integrating emission from clouds with a broad range of density and ionizing flux, LOC models include the optimally emitting clouds for all lines of interest. The integration is typically done over power-law distributions of $n_{H}$ and $\Phi_{H}$ (or alternatively radius from the continuum source) \citep{baldwin95,korista00}. For our LOC models we integrated over an entire grid with a function of the form
\begin{equation}
\int\int{AF(n_H,\Phi_H)n_H^{\beta_n}\Phi_H^{\beta_\Phi}\,d(\log{n_H})d(\log{\Phi_H})} \; ,\label{eqn_loc}
\end{equation}
where $F$ is the model flux of the line and $A$, $\beta_n$, and $\beta_\Phi$ are free parameters of the model. Best-fit values of the three free parameters were determined using MPFIT.  

As for the single-component models, we fit an LOC model to each individual AGN spectrum for every set of spectral indices. Only one spectrum, SDSS~J124154.02+572107.3, was well fit with the \Nthree $\lambda$991 line included. It was successfully fit for four different sets of spectral indices. Best-fit LOC models for the other individual spectra predicted \Nthree fluxes between 22\% (HS~1102+3441) and 73\% (HE~0226-4110) of the measured values.
Fitting LOC models with \Nthree excluded produced good fits to an additional three individual spectra, with HE~0226-4110, RXJ~2154.1-4414, and SDSS~J132222.68+464535.2 having good fits for one, two, and three pairs of spectral indices, respectively. For HS~1102+3441, the equivalent widths of all emission lines were underpredicted by the LOC models. As can be seen in Figure \ref{fig3}, the equivalent widths for HS~1102+3441 are significantly higher than those of the other lines. This AGN likely requires a covering factor larger than 40\% to account for the strength of its BLR emission. LOC models were also fit to the composite spectrum for both the full set of 20 lines and for the five lines that were fit in the individual spectra. No models with all 20 lines provided good fits, with typical $\chi^2$ values $>500$. There was one good fit to the composite spectrum with only five lines. Table \ref{tab4} summarizes the properties of all LOC models with $\chi^2$ values corresponding to a value of $\le0.9$ for the $\chi^2$ cumulative distribution ($\chi^2 < 2.7$ for 1 dof, $\chi^2 < 4.6$ for 2 dof, and $\chi^2 < 6.3$ for 3 dof).

\begin{figure}
\plotone{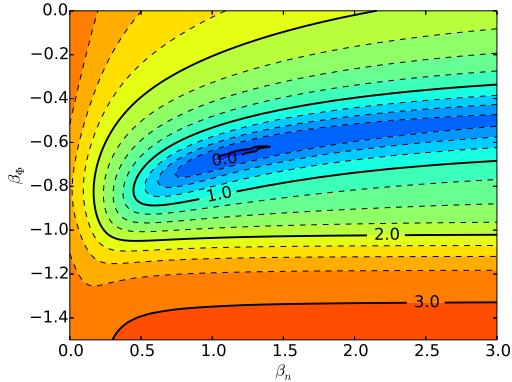}
\caption{Contour plot of $\log{\chi^2}$ for the LOC model fit to HE 0226-4110 with spectral indices $\alpha_{UV} = -1.0$ and $\alpha_{\text{trans}} = -1.5$. The best-fit values for this model are $\beta_n = 1.161\pm0.912$ and $\beta_\Phi = -0.653\pm0.173$.\label{LOC}}
\end{figure}

Table \ref{tab4} shows large differences in the best-fit values of $\beta_n$ between the spectra. Previous studies using LOC models have typically used density distributions that scale as $n^{-1}\,dn$, which corresponds to $\beta_n = 0$ in our models \citep{baldwin95,korista00}. This is consistent with most of our results for RXJ~2154.1-4414 and SDSS~J132222.68+464535.2. However, our fits for HE~0226-4110 and SDSS~J124154.02+572107.3 have distributions with much greater contributions from high-density clouds. The contour plot of $\log{\chi^2}$ for the HE~0226-4110 fit (Figure \ref{LOC}) makes it clear that this difference is not simply due to statistical uncertainties in the fit parameters. The primary difference between these two sets of spectra is that HE~0226-4110 and SDSS~J124154.02+572107.3 include EUV lines shortward of 900\AA, while the \Nfive spectra do not. As discussed in Section \ref{EW}, higher $T_\text{exc}$ lines are produced preferentially in high $n_H$ and $\Phi_H$ clouds. It is likely that the flux from these EUV lines in HE~0226-4110 and SDSS~J124154.02+572107.3 is driving the large values of $\beta_n$. The value of $\beta_n$ also depends on the UV spectral index of the model. In general, harder UV spectral indices correspond to larger $\beta_n$. This is particularly true for the main spectra, where $\beta_n$ is negative or consistent with zero for $\alpha_{UV} = -1.5$, but often greater than one for $\alpha_{UV} = -1.0$ or $-0.5$. The trend is less pronounced in the \Nfive spectra, with $-0.4 < \lvert\beta_n\rvert < 0.3$ in all but one case, but it is still present. This relationship makes physical sense, as more ionizing UV photons means that higher density clouds are required to achieve the same photoionization structure.

The limits of integration in Equation \ref{eqn_loc} also affect the parameters of our fits. All of the LOC models in Table \ref{tab4} used the full \Cloudy grids, with $7 \le \log n_H (\text{cm}^{-3}) \le 14$ and $17 \le \log \Phi_H (\text{cm}^{-2}\text{ s}^{-1}) \le 24$. We fit LOC models with different limits of integration to SDSS~J124154.02+572107.3 and SDSS~J132222.68+464535.2, to investigate the effects on our main and \Nfive spectra, respectively. For SDSS~J124154.02+572107.3, increasing the lower limit on $n_H$ to $10^8$ cm$^{-3}$ or the limit on $\Phi_H$ to $10^{18}$ cm$^{-2}$ s$^{-1}$ had almost no effect. All good fits for the full \Cloudy grids remained good fits, and the best-fit values of $\beta_n$ and $\beta_\Phi$ were consistent within their errorbars. In contrast, decreasing the upper limit on $n_H$ to $10^{13}$ cm$^{-3}$ significantly worsened the fits for most pairs of spectral indices, producing no good fits. In addition, the best-fit values of $\beta_H$ increased, particularly for the models with harder UV spectral indices. For example, the grid with $\alpha_{UV} = -1.0$ and $\alpha_\text{trans} = -1.5$ had the best-fit value of $\beta_H$ increase from 1.299 to 3.592. Reducing the upper bound on $\Phi_H$ to $10^{23}$ cm$^{-2}$ s$^{-1}$ produced a similar worsening of the LOC fits. Here, the best-fit values of $\beta_\Phi$ increased from $\sim-1$ to values between 0 and 2 for $\alpha_{UV} = -1.0$ or $-0.5$, while $\beta_H$ decreased. These changes in the LOC parameters are consistent with expectations. As mentioned before, the LOC fits to the spectra with EUV lines require large contributions from high $n_H$ and $\Phi_H$ clouds, so decreasing the upper limits of integration in our models requires even large contributions from the remaining clouds to compensate.

For SDSS~J132222.68+464535.2, reducing the upper limits on $n_H$ to $10^{13}$ cm$^{-3}$ or $\Phi_H$ to $10^{23}$ cm$^{-2}$ s$^{-1}$, has little impact on the LOC fits. The best-fit $\chi^2$ values change by $\sim10\%$, so the fit for $\alpha_{UV} = -0.5$ and $\alpha_\text{trans} = -2.0$, which was borderline with the full grid, is no longer a good fit. The values of $\beta_n$ and $\beta_\Phi$ are consistent with the fits from the full grid to within their errorbars. The same is true when the lower limit on $n_H$ is increased to $10^8$, although in this case, only one of the good fits from the full grid remains. In contrast, when the lower limit on $\Phi_H$ is increased to $10^{18}$ cm$^{-2}$ s$^{-1}$, the $\chi^2$ values of many of the previously poor fits with the full grid are reduced significantly, although not by enough to provide additional good fits. The best-fit values of $\beta_\Phi$ are changed from being clustered tightly between $-1.1$ and $-0.9$ to ranging from $-2.7$ to $-1.0$, with typical values around $-1.3$. The values of $\beta_n$ are not changed significantly. Figure \ref{fig5} shows that the peak Ly$\alpha$ emission occurs for $\Phi_H < 10^{18}$ cm$^{-2}$ s$^{-1}$, so the change in $\beta_\Phi$ is likely caused by a need for more low ionizing flux clouds to compensate for the increased lower limit. The differences between the effects of the limits of integration on SDSS~J124154.02+572107.3 and SDSS~J132222.68+464535.2 further emphasizes the importance of including EUV emission lines in BLR studies.

\begin{deluxetable*}{lrrrcrr}
\tablewidth{0pt}
\tablecaption{LOC Fits to Observed Spectra\label{tab4}} 
\tablehead{\colhead{Spectrum} & \colhead{$\alpha_{UV}$\tablenotemark{a}} & \colhead{$\alpha_{\text{trans}}$\tablenotemark{a}} & \colhead{$\chi^2$} & \colhead{dof} & \colhead{$\beta_n$} & \colhead{$\beta_\Phi$}}
\startdata
HE 0226-4110 (no \Nthree) & $-1.0$ & $-1.5$ & $0.95$ & $1$ & $1.161\pm0.912$ & $-0.653\pm0.173$ \\
SDSS J124154.02+572107.3 & $-0.5$ & $-1.5$ & $1.09$ & $2$ & $2.086\pm1.816$ & $-0.702\pm0.075$ \\
\nodata & $-1.0$ & $-1.5$ & $0.96$ & $2$ & $1.282\pm0.675$ & $-0.881\pm0.026$ \\
\nodata & $-1.0$ & $-2.0$ & $0.89$ & $2$ & $2.021\pm0.674$ & $-0.798\pm0.065$ \\
\nodata & $-1.0$ & $-2.5$ & $1.11$ & $2$ & $1.653\pm1.527$ & $-0.729\pm0.057$ \\
RXJ 2154.1-4414 (no \Nthree) & $-0.5$ & $-2.0$ & $0.68$ & $2$ & $0.033\pm0.049$ & $-0.985\pm0.019$ \\
\nodata & $-1.0$ & $-1.5$ & $4.21$ & $2$ & $-0.016\pm0.041$ & $-1.080\pm0.019$ \\
SDSS J132222.68+464535.2 (no \Nthree) & $-0.5$ & $-2.0$ & $4.51$ & $2$ & $0.027\pm0.055$ & $-0.976\pm0.016$ \\
\nodata & $-1.0$ & $-1.5$ & $1.64$ & $2$ & $0.002\pm0.046$ & $-1.061\pm0.018$ \\
\nodata & $-1.5$ & $-1.5$ & $2.53$ & $2$ & $-0.235\pm0.047$ & $-1.149\pm0.028$ \\
Composite & $-0.5$ & $-1.5$ & $5.24$ & $3$ & $0.133\pm0.034$ & $-0.739\pm0.032$ \\ 
\enddata
\tablenotetext{a}{Spectral indices with respect to frequency ($F_\nu \propto \nu^{\alpha_\nu}$)}
\end{deluxetable*} 

\section{DISCUSSION}\label{discuss}

Overall, our attempts to fit single-component \Cloudy models to the measured EUV line fluxes were unsuccessful. Only two of the individual spectra provided adequate fits for any set of spectral indices, and only after the \Nthree $\lambda$991 emission line was excluded from the fit. This result is not particularly surprising. As Figures \ref{fig5} and \ref{fig6} show, each emission line originates from clouds with different physical properties and locations within the BLR. Any model that seeks to fit multiple emission lines needs to include emission from a distribution of clouds in $n_H$ and $\Phi_H$. Our LOC models do this, and as a result they are much more effective, producing good fits for four of the six individual spectra as well as the set of five EUV lines from the composite spectrum. Previous studies have successfully fit LOC models to both individual and composite spectra, but using UV lines longward of Ly$\alpha$ \citep{baldwin95,korista00}. Our results extend the validity of this model into the EUV. One important caveat for all of the following analysis is that it is only valid if our \Cloudy grids and LOC models provide an accurate picture of the BLR. Alternative models include ones with different velocity profiles such as disk winds \citep{gaskell82}, and different geometries such as the ``bowl-shaped'' BLR in \citet{goad12}.

While our LOC models successfully fit the subset of five EUV lines taken from the composite spectrum, they provide poor fits to the composite as a whole. It is unclear whether this is a result specific to the composite or represents a general issue with extending LOC models over a large range of excitation energies. The spectral range of the composite is achieved by combining spectra from AGN with redshifts that vary systematically across the spectrum. If the BLR properties of AGN or their host environments vary significantly with redshift, then any attempt at modeling a composite spectrum as a single object will be invalid. It is also possible that LOC models are in general not valid over large ranges in excitation energy. Our LOC fits to the main spectra require larger contributions from high density clouds than the \Nfive spectra do, reflected in a larger $\beta_n$, even though they share many of the same spectral lines. Any spectrum covering a broad UV wavelength range may require more than a simple power law in density to accurately models its emission.

\newpage
\subsection{UV Spectral Indices}
The qualities of our different LOC fits can be used to constrain the shape of the incident SED and compare it to our observed continua. \citet{korista97_2} found evidence for harder intrinsic UV spectra by comparing \Cloudy models of \Hetwo $\lambda$1640 emission to observations of Mrk 335 and a composite spectrum from \citet{zheng97}. Our results do not show a similar preference for harder intrinsic spectra. All four of the individual spectra with good LOC fits have model $\alpha_{UV}$ values consistent with the observations. The best fit LOC model for HE~0226-4110 has $\alpha_{UV} = -1.0$ compared to the observed value of $-0.930$, while for SDSS~J124154.02+572107.3, the best-fit UV indices are all $-0.5$ or $-1.0$, consistent with the observed EUV spectral index of $-0.598$. For the \Nfive spectra, the fits to RXJ~2154.1-4414 have $\alpha_{UV} = -0.5$ or $-1.0$, compared to the observed value of $-0.676$. The good fits to SDSS~J132222.68+464535.2 include all three $\alpha_{UV}$ values, so these are also consistent with the observed $-0.646$. In contrast, the UV spectral index for the fit to the composite has $\alpha_{UV} = -0.5$, which is significantly harder than the observed EUV spectral index of $-1.401$, and also harder than the FUV index of $-0.828$. However, as discussed above, the validity of fitting LOC models to a composite spectrum is uncertain. In addition, $\alpha_{UV} = -0.5$ is still well within the range of spectral indices found for individual objects. One of the two individual spectra without a good LOC fit, HE~0238-1904, has an observed EUV spectral index ($\alpha_{UV} = -1.902$) significantly softer than in any of our model grids, which may explain our inability to obtain a successful fit.

In their analysis of intrinsic extinction in AGN, \citet{gaskell04} argue that most of the observed variation between AGN continua can be explained by extinction and that AGN have an intrinsic UV spectral index of $\sim-0.45$. Three of the four spectra with good LOC fits, as well as the composite, do have fits with $\alpha_{UV} = -0.5$. However, for the individual spectra this UV index is also consistent with the measured spectral indices, which have not been corrected for intrinsic reddening. All four individual spectra have good fits for $\alpha_{UV} = -1.0$, so there is no preference in our models for a value of $-0.5$ over this softer slope. Combined with the coarse resolution of our spectral index grid, there is no clear evidence that the intrinsic UV indices of our AGN are harder than their observed values.

\subsection{\Nthree $\lambda$991 Flux}\label{Nthree}

A common theme throughout this paper has been the differences between \Nthree $\lambda$991 and the rest of our emission lines. The measured equivalent widths for \Nthree show larger variation between the different AGN than any of the other lines, with a total spread of over an order of magnitude, and all of the \Nthree equivalent widths are above that of the composite. In addition, our model fits underpredict the observed \Nthree fluxes by as much as a factor of five. This is particularly bad for HE~0226-4110, RXJ~2154.1-4414, and SDSS~J132222.68+464535.2, which can only be successfully fit when \Nthree is excluded from the models. 

One possible explanation for this is that our fits to the \Nthree emission include more flux from other lines than is accounted for by our deblending analysis. In Figure \ref{fig2} there is a significant tail in the \Nthree fit extending past 1000 \AA{}. It is possible that much of this flux actually originates from the \Osix doublet or Ly$\beta$. If there is a highly blueshifted portion of the \Osix emission, perhaps from a wind, the wavelength restrictions placed on our fits would not allow it to be properly included in the \Osix line. The plausibility of this scenario can be investigated with some simple estimates. For there to be a significant effect on the \Nthree measurement, $\gtrsim5\%$ of the \Osix flux would need to be shortward of 1000 \AA, where it would be attributed to \Nthree in the fits. \citet{corbin90} find that the largest relative blueshifts between emission lines are $\sim4000$ km s$^{-1}$, which corresponds to a blueshift of 14 \AA. This shifts the center of the \Osix doublet to $\sim1020$ \AA. From our fits to the \Osix emission, typically $\sim50\%$ of the flux is in a broad ($\sigma = $10-15 \AA) component. For $\sigma = 15$ \AA, this leaves $5\%$ of the \Osix flux below 1000 \AA, as required. So, the blueshifts required to explain the excess \Nthree in this way are strong, but still consistent with observations.

Another cause of the strong \Nthree emission could be increased metallicity or nitrogen abundance above the solar values used in our \Cloudy grids. The QSO chemical evolution models in \citet{hamann93} predict increased nitrogen abundance relative to other elements with increasing metallicity. Both \citet{baldwin95} and \citet{korista00} find that \Cloudy models underpredict \Nfive $\lambda$1240 emission for solar abundances. However, the data from our \Nfive spectra makes this explanation less likely. If increased nitrogen abundance was the reason for the higher than predicted \Nthree fluxes, then we would expect the \Nfive emission to be higher than predicted as well. Instead, our \Cloudy models underpredict the \Nthree emission while accurately predicting that of \Nfive.

It is not possible to rule out elevated nitrogen abundances with this study. Fully testing this would require \Cloudy models with variations in both nitrogen abundance and overall metallicity, which we lack the necessary degrees of freedom to test. We did, however, conduct one simple test by running a \Cloudy grid with $\alpha_{UV} = -1.0$ and $\alpha_\text{trans} = -1.5$ with a nitrogen abundance of five times solar. As can be seen in Table \ref{tab4}, the grid with these spectral indices and solar abundances provided good fits to four individual spectra, although three of these required the \Nthree flux to be omitted. We fit LOC models to the new grid with elevated nitrogen for all six individual spectra, with the \Nthree line included in all fits. There was one good fit, with HE~0226-4110 having $\chi^2 = 2.38$ for 2 dof, as compared to $\chi^2 = 8.64$ for the grid with solar abundances. However, none of the other five spectra produced good LOC fits, and for SDSS~J132222.68+464535.2, RXJ~2154.1-4414, and SDSS~J132222.68+464535.2, the fits with elevated nitrogen abundances were worse than the ones with solar values. Given this data, the most likely explanation for the \Nthree $\lambda$991 flux is deblending errors or other properties of the line fits.

\section{CONCLUSIONS}\label{sum}

In this paper we measure EUV emission line fluxes from single-epoch spectra of four individual AGN with $0.49 \le z \le 0.64$, two AGN with $0.32 \le z \le 0.40$, and a composite of 159 AGN spectra from \citet{stevans14}. We then use grids of \Cloudy photoionization model calculations to fit BLR models to each of the spectra for a variety of incident IR to UV and 4 - 30 Ryd continuum spectral indices. Our results can be summarized as follows:

\begin{enumerate}
\item The peak emission from EUV lines of \Nfour, \Ofour, \Othree, and \Otwo comes from clouds with electron temperatures above $20\%$ of the excitation temperature of the line, favoring clouds with high density and incident ionizing flux. Typical electron temperatures for this emission are between 37,000~K and 55,000~K.
\item Single-component \Cloudy models only provide good fits for two of the six individual spectra, and only when the \Nthree $\lambda$991 flux is excluded from the models.
\item LOC models provide good fits to four of our six individual AGN, with three having fits for multiple pairs of incident spectral indices.
\item Only one of the individual spectra, HS~1102+3441, clearly requires a covering factor greater than our adopted value of 40\%.
\item The EUV lines in the composite spectrum are well fit by locally optimally emitting cloud (LOC) models, but the full spectral range of the composite is not.
\item The UV spectral indices preferred by our LOC models are consistent with the EUV continuum fit to our observed spectra.
\item The observed flux in \Nthree $\lambda$991 is consistently underpredicted by our models. This is likely due to the nature of our line fits rather than elevated nitrogen abundances.
\end{enumerate}

\acknowledgements
We thank Charles Danforth and Matthew Stevans for providing coadditions of the COS spectra. Matthew Stevans additionally provided continuum fits for the individual spectra and the composites. Evan Tilton provided helpful comments on this project. This work was supported by NASA grants NNX07-AG77G and NNX08-AC14G. JMS thanks the Institute of Astronomy at Cambridge University for their stimulating scientific environment and support through the Sackler Visitor Program.

\bibliographystyle{apj}
\bibliography{ref}

\begin{thebibliography}{}
\expandafter\ifx\csname natexlab\endcsname\relax\def\natexlab#1{#1}\fi

\bibitem[{{Arav} {et~al.}(2008){Arav}, {Moe}, {Costantini}, {Korista}, {Benn},
  \& {Ellison}}]{arav08}
{Arav}, N., {Moe}, M., {Costantini}, E., {et~al.} 2008, \apj, 681, 954

\bibitem[{{Bajtlik} {et~al.}(1988){Bajtlik}, {Duncan}, \&
  {Ostriker}}]{bajtlik88}
{Bajtlik}, S., {Duncan}, R.~C., \& {Ostriker}, J.~P. 1988, \apj, 327, 570

\bibitem[{{Baldwin} {et~al.}(1995){Baldwin}, {Ferland}, {Korista}, \&
  {Verner}}]{baldwin95}
{Baldwin}, J., {Ferland}, G., {Korista}, K., \& {Verner}, D. 1995, \apjl, 455,
  L119

\bibitem[{{Baldwin}(1997)}]{baldwin97}
{Baldwin}, J.~A. 1997, in Astronomical Society of the Pacific Conference
  Series, Vol. 113, IAU Colloq. 159: Emission Lines in Active Galaxies: New
  Methods and Techniques, ed. B.~M. {Peterson}, F.-Z. {Cheng}, \& A.~S.
  {Wilson}, 80

\bibitem[{{Blandford} \& {McKee}(1982)}]{blandford82}
{Blandford}, R.~D., \& {McKee}, C.~F. 1982, \apj, 255, 419

\bibitem[{{Carswell} {et~al.}(1987){Carswell}, {Webb}, {Baldwin}, \&
  {Atwood}}]{carswell87}
{Carswell}, R.~F., {Webb}, J.~K., {Baldwin}, J.~A., \& {Atwood}, B. 1987, \apj,
  319, 709

\bibitem[{{Corbin}(1990)}]{corbin90}
{Corbin}, M.~R. 1990, \apj, 357, 346

\bibitem[{{Croton} {et~al.}(2006){Croton}, {Springel}, {White}, {De Lucia},
  {Frenk}, {Gao}, {Jenkins}, {Kauffmann}, {Navarro}, \& {Yoshida}}]{croton06}
{Croton}, D.~J., {Springel}, V., {White}, S.~D.~M., {et~al.} 2006, \mnras, 365,
  11

\bibitem[{{Danforth} {et~al.}(2010){Danforth}, {Keeney}, {Stocke}, {Shull}, \&
  {Yao}}]{coadd}
{Danforth}, C.~W., {Keeney}, B.~A., {Stocke}, J.~T., {Shull}, J.~M., \& {Yao},
  Y. 2010, \apj, 720, 976

\bibitem[{{Danforth} \& {Shull}(2008)}]{danforth08}
{Danforth}, C.~W., \& {Shull}, J.~M. 2008, \apj, 679, 194

\bibitem[{{Di Matteo} {et~al.}(2005){Di Matteo}, {Springel}, \&
  {Hernquist}}]{dimatteo05}
{Di Matteo}, T., {Springel}, V., \& {Hernquist}, L. 2005, \nat, 433, 604

\bibitem[{{Dunn} {et~al.}(2007){Dunn}, {Crenshaw}, {Kraemer}, \&
  {Gabel}}]{dunn07}
{Dunn}, J.~P., {Crenshaw}, D.~M., {Kraemer}, S.~B., \& {Gabel}, J.~R. 2007,
  \aj, 134, 1061

\bibitem[{{Dunn} {et~al.}(2010){Dunn}, {Bautista}, {Arav}, {Moe}, {Korista},
  {Costantini}, {Benn}, {Ellison}, \& {Edmonds}}]{dunn10}
{Dunn}, J.~P., {Bautista}, M., {Arav}, N., {et~al.} 2010, \apj, 709, 611

\bibitem[{{Ferland} {et~al.}(2013){Ferland}, {Porter}, {van Hoof}, {Williams},
  {Abel}, {Lykins}, {Shaw}, {Henney}, \& {Stancil}}]{ferland13}
{Ferland}, G.~J., {Porter}, R.~L., {van Hoof}, P.~A.~M., {et~al.} 2013, Rev.
  Mexicana Astron. Astrofis., 49, 137

\bibitem[{{Ferrarese} \& {Merritt}(2000)}]{ferrarese00}
{Ferrarese}, L., \& {Merritt}, D. 2000, \apjl, 539, L9

\bibitem[{{Fitzpatrick}(1999)}]{fitzpatrick99}
{Fitzpatrick}, E.~L. 1999, \pasp, 111, 63

\bibitem[{{Gaskell}(1982)}]{gaskell82}
{Gaskell}, C.~M. 1982, \apj, 263, 79

\bibitem[{{Gaskell}(2009)}]{gaskell09}
---. 2009, New A Rev., 53, 140

\bibitem[{{Gaskell} \& {Goosmann}(2013)}]{gaskell13}
{Gaskell}, C.~M., \& {Goosmann}, R.~W. 2013, \apj, 769, 30

\bibitem[{{Gaskell} {et~al.}(2004){Gaskell}, {Goosmann}, {Antonucci}, \&
  {Whysong}}]{gaskell04}
{Gaskell}, C.~M., {Goosmann}, R.~W., {Antonucci}, R.~R.~J., \& {Whysong}, D.~H.
  2004, \apj, 616, 147

\bibitem[{{Gaskell} {et~al.}(2007){Gaskell}, {Klimek}, \&
  {Nazarova}}]{gaskell07}
{Gaskell}, C.~M., {Klimek}, E.~S., \& {Nazarova}, L.~S. 2007, ArXiv e-prints,
  arXiv:0711.1025

\bibitem[{{Gebhardt} {et~al.}(2000){Gebhardt}, {Bender}, {Bower}, {Dressler},
  {Faber}, {Filippenko}, {Green}, {Grillmair}, {Ho}, {Kormendy}, {Lauer},
  {Magorrian}, {Pinkney}, {Richstone}, \& {Tremaine}}]{gebhardt00}
{Gebhardt}, K., {Bender}, R., {Bower}, G., {et~al.} 2000, \apjl, 539, L13

\bibitem[{{Goad} {et~al.}(2012){Goad}, {Korista}, \& {Ruff}}]{goad12}
{Goad}, M.~R., {Korista}, K.~T., \& {Ruff}, A.~J. 2012, \mnras, 426, 3086

\bibitem[{{Gon{\c c}alves} {et~al.}(2008){Gon{\c c}alves}, {Steidel}, \&
  {Pettini}}]{goncalves08}
{Gon{\c c}alves}, T.~S., {Steidel}, C.~C., \& {Pettini}, M. 2008, \apj, 676,
  816

\bibitem[{{Green} {et~al.}(2012){Green}, {Froning}, {Osterman}, {Ebbets},
  {Heap}, {Leitherer}, {Linsky}, {Savage}, {Sembach}, {Shull}, {Siegmund},
  {Snow}, {Spencer}, {Stern}, {Stocke}, {Welsh}, {B{\'e}land}, {Burgh},
  {Danforth}, {France}, {Keeney}, {McPhate}, {Penton}, {Andrews},
  {Brownsberger}, {Morse}, \& {Wilkinson}}]{green12}
{Green}, J.~C., {Froning}, C.~S., {Osterman}, S., {et~al.} 2012, \apj, 744, 60

\bibitem[{{G{\"u}ltekin} {et~al.}(2009){G{\"u}ltekin}, {Richstone}, {Gebhardt},
  {Lauer}, {Tremaine}, {Aller}, {Bender}, {Dressler}, {Faber}, {Filippenko},
  {Green}, {Ho}, {Kormendy}, {Magorrian}, {Pinkney}, \& {Siopis}}]{gultekin09}
{G{\"u}ltekin}, K., {Richstone}, D.~O., {Gebhardt}, K., {et~al.} 2009, \apj,
  698, 198

\bibitem[{{Haardt} \& {Madau}(2012)}]{haardt12}
{Haardt}, F., \& {Madau}, P. 2012, \apj, 746, 125

\bibitem[{{Hamann} \& {Ferland}(1993)}]{hamann93}
{Hamann}, F., \& {Ferland}, G. 1993, \apj, 418, 11

\bibitem[{{Hopkins} {et~al.}(2006){Hopkins}, {Hernquist}, {Cox}, {Di Matteo},
  {Robertson}, \& {Springel}}]{hopkins06}
{Hopkins}, P.~F., {Hernquist}, L., {Cox}, T.~J., {et~al.} 2006, \apjs, 163, 1

\bibitem[{{Keeney} {et~al.}(2012){Keeney}, {Danforth}, {Stocke}, {France}, \&
  {Green}}]{coadd2}
{Keeney}, B.~A., {Danforth}, C.~W., {Stocke}, J.~T., {France}, K., \& {Green},
  J.~C. 2012, \pasp, 124, 830

\bibitem[{{Korista} {et~al.}(1997{\natexlab{a}}){Korista}, {Baldwin},
  {Ferland}, \& {Verner}}]{korista97_1}
{Korista}, K., {Baldwin}, J., {Ferland}, G., \& {Verner}, D.
  1997{\natexlab{a}}, \apjs, 108, 401

\bibitem[{{Korista} {et~al.}(1997{\natexlab{b}}){Korista}, {Ferland}, \&
  {Baldwin}}]{korista97_2}
{Korista}, K., {Ferland}, G., \& {Baldwin}, J. 1997{\natexlab{b}}, \apj, 487,
  555

\bibitem[{{Korista} \& {Goad}(2000)}]{korista00}
{Korista}, K.~T., \& {Goad}, M.~R. 2000, \apj, 536, 284

\bibitem[{{Kormendy}(1993)}]{kormendy93}
{Kormendy}, J. 1993, in The Nearest Active Galaxies, ed. J.~{Beckman},
  L.~{Colina}, \& H.~{Netzer}, 197--218

\bibitem[{{Kormendy} \& {Ho}(2013)}]{kormendy13}
{Kormendy}, J., \& {Ho}, L.~C. 2013, \araa, 51, 511

\bibitem[{{Kormendy} \& {Richstone}(1995)}]{kormendy95}
{Kormendy}, J., \& {Richstone}, D. 1995, \araa, 33, 581

\bibitem[{{Lehner} {et~al.}(2007){Lehner}, {Savage}, {Richter}, {Sembach},
  {Tripp}, \& {Wakker}}]{lehner07}
{Lehner}, N., {Savage}, B.~D., {Richter}, P., {et~al.} 2007, \apj, 658, 680

\bibitem[{{Markwardt}(2009)}]{mpfit}
{Markwardt}, C.~B. 2009, in Astronomical Society of the Pacific Conference
  Series, Vol. 411, Astronomical Data Analysis Software and Systems XVIII, ed.
  D.~A. {Bohlender}, D.~{Durand}, \& P.~{Dowler}, 251

\bibitem[{{Mathews} \& {Ferland}(1987)}]{mathews87}
{Mathews}, W.~G., \& {Ferland}, G.~J. 1987, \apj, 323, 456

\bibitem[{{McNamara} {et~al.}(2005){McNamara}, {Nulsen}, {Wise}, {Rafferty},
  {Carilli}, {Sarazin}, \& {Blanton}}]{mcnamara05}
{McNamara}, B.~R., {Nulsen}, P.~E.~J., {Wise}, M.~W., {et~al.} 2005, \nat, 433,
  45

\bibitem[{{Oppenheimer} \& {Schaye}(2013)}]{oppenheimer13}
{Oppenheimer}, B.~D., \& {Schaye}, J. 2013, \mnras, 434, 1063

\bibitem[{{Osterman} {et~al.}(2011){Osterman}, {Green}, {Froning},
  {B{\'e}land}, {Burgh}, {France}, {Penton}, {Delker}, {Ebbets}, {Sahnow},
  {Bacinski}, {Kimble}, {Andrews}, {Wilkinson}, {McPhate}, {Siegmund}, {Ake},
  {Aloisi}, {Biagetti}, {Diaz}, {Dixon}, {Friedman}, {Ghavamian}, {Goudfrooij},
  {Hartig}, {Keyes}, {Lennon}, {Massa}, {Niemi}, {Oliveira}, {Osten},
  {Proffitt}, {Smith}, \& {Soderblom}}]{osterman11}
{Osterman}, S., {Green}, J., {Froning}, C., {et~al.} 2011, \apss, 335, 257

\bibitem[{{Peek} \& {Schiminovich}(2013)}]{peek13}
{Peek}, J.~E.~G., \& {Schiminovich}, D. 2013, \apj, 771, 68

\bibitem[{{Penton} {et~al.}(2000){Penton}, {Shull}, \& {Stocke}}]{penton00}
{Penton}, S.~V., {Shull}, J.~M., \& {Stocke}, J.~T. 2000, \apj, 544, 150

\bibitem[{{Penton} {et~al.}(2004){Penton}, {Stocke}, \& {Shull}}]{penton04}
{Penton}, S.~V., {Stocke}, J.~T., \& {Shull}, J.~M. 2004, \apjs, 152, 29

\bibitem[{{Peterson}(1993)}]{peterson93}
{Peterson}, B.~M. 1993, \pasp, 105, 247

\bibitem[{{Reimers} {et~al.}(1997){Reimers}, {Kohler}, {Wisotzki}, {Groote},
  {Rodriguez-Pascual}, \& {Wamsteker}}]{reimers97}
{Reimers}, D., {Kohler}, S., {Wisotzki}, L., {et~al.} 1997, \aap, 327, 890

\bibitem[{{Schlafly} \& {Finkbeiner}(2011)}]{dustmap}
{Schlafly}, E.~F., \& {Finkbeiner}, D.~P. 2011, \apj, 737, 103

\bibitem[{{Schlegel} {et~al.}(1998){Schlegel}, {Finkbeiner}, \&
  {Davis}}]{schlegel98}
{Schlegel}, D.~J., {Finkbeiner}, D.~P., \& {Davis}, M. 1998, \apj, 500, 525

\bibitem[{{Scott} {et~al.}(2004){Scott}, {Kriss}, {Brotherton}, {Green},
  {Hutchings}, {Shull}, \& {Zheng}}]{scott04}
{Scott}, J.~E., {Kriss}, G.~A., {Brotherton}, M., {et~al.} 2004, \apj, 615, 135

\bibitem[{{Shull} {et~al.}(2012{\natexlab{a}}){Shull}, {Smith}, \&
  {Danforth}}]{shull12}
{Shull}, J.~M., {Smith}, B.~D., \& {Danforth}, C.~W. 2012{\natexlab{a}}, \apj,
  759, 23

\bibitem[{{Shull} {et~al.}(2012{\natexlab{b}}){Shull}, {Stevans}, \&
  {Danforth}}]{composite1}
{Shull}, J.~M., {Stevans}, M., \& {Danforth}, C.~W. 2012{\natexlab{b}}, \apj,
  752, 162

\bibitem[{{Shull} {et~al.}(2004){Shull}, {Tumlinson}, {Giroux}, {Kriss}, \&
  {Reimers}}]{shull04}
{Shull}, J.~M., {Tumlinson}, J., {Giroux}, M.~L., {Kriss}, G.~A., \& {Reimers},
  D. 2004, \apj, 600, 570

\bibitem[{{Silk} \& {Norman}(2009)}]{silk09}
{Silk}, J., \& {Norman}, C. 2009, \apj, 700, 262

\bibitem[{{Silk} \& {Rees}(1998)}]{silk98}
{Silk}, J., \& {Rees}, M.~J. 1998, \aap, 331, L1

\bibitem[{{Stevans} {et~al.}(2014){Stevans}, {Shull}, {Danforth}, \&
  {Tilton}}]{stevans14}
{Stevans}, M.~L., {Shull}, J.~M., {Danforth}, C.~W., \& {Tilton}, E.~M. 2014,
  \apj, submitted

\bibitem[{{Telfer} {et~al.}(2002){Telfer}, {Zheng}, {Kriss}, \&
  {Davidsen}}]{telfer02}
{Telfer}, R.~C., {Zheng}, W., {Kriss}, G.~A., \& {Davidsen}, A.~F. 2002, \apj,
  565, 773

\bibitem[{{Thom} \& {Chen}(2008)}]{thom08}
{Thom}, C., \& {Chen}, H.-W. 2008, \apj, 683, 22

\bibitem[{{Tilton} {et~al.}(2012){Tilton}, {Danforth}, {Shull}, \&
  {Ross}}]{tilton12}
{Tilton}, E.~M., {Danforth}, C.~W., {Shull}, J.~M., \& {Ross}, T.~L. 2012,
  \apj, 759, 112

\bibitem[{{Tripp} {et~al.}(2008){Tripp}, {Sembach}, {Bowen}, {Savage},
  {Jenkins}, {Lehner}, \& {Richter}}]{tripp08}
{Tripp}, T.~M., {Sembach}, K.~R., {Bowen}, D.~V., {et~al.} 2008, \apjs, 177, 39

\bibitem[{{Vestergaard} \& {Peterson}(2006)}]{vestergaard06}
{Vestergaard}, M., \& {Peterson}, B.~M. 2006, \apj, 641, 689

\bibitem[{{Wang} {et~al.}(2012){Wang}, {Ferland}, {Hu}, {Wang}, \&
  {Du}}]{wang12}
{Wang}, Y., {Ferland}, G.~J., {Hu}, C., {Wang}, J.-M., \& {Du}, P. 2012,
  \mnras, 424, 2255

\bibitem[{{Zheng} {et~al.}(1997){Zheng}, {Kriss}, {Telfer}, {Grimes}, \&
  {Davidsen}}]{zheng97}
{Zheng}, W., {Kriss}, G.~A., {Telfer}, R.~C., {Grimes}, J.~P., \& {Davidsen},
  A.~F. 1997, \apj, 475, 469

\end{thebibliography}

\clearpage
\begin{turnpage}
\begin{deluxetable*}{llrrrrrrr}
\tablewidth{0pt}
\tablecaption{Emission Line Fluxes and Errors\label{tab2}}
\tablehead{\colhead{AGN} & \colhead{Line} & \colhead{Flux\tablenotemark{a}} & \colhead{Stat. Error\tablenotemark{b}} & \colhead{Blend Error\tablenotemark{b}} & \colhead{E(B-V) Error\tablenotemark{b}} & \colhead{$R_V$ Error\tablenotemark{b}} & \colhead{Total Error\tablenotemark{b}} & \colhead{EW (\AA{})}}
\startdata
HE 0226-4110 & \Osix & $4.72\e{-13}$ & $\pm\phn0.5\%$ & $\pm\phn7.5\%$ & $\pm1.5\%$ & $\pm0.3\%$ & $\pm\phn7.7\%$ & 25.21 \\
\nodata & \Nthree & $1.18\e{-13}$ & $\pm\phn4.2\%$ & $\pm15.0\%$ & $\pm1.5\%$ & $\pm0.4\%$ & $\pm16.2\%$ & 6.02\\
\nodata & \Cthree & $1.04\e{-13}$ & $\pm\phn5.4\%$ & $\pm15.0\%$ & $\pm1.4\%$ & $\pm0.4\%$ & $\pm16.1\%$ & 5.22\\
\nodata & \Otwo + \Othree & $5.13\e{-14}$ & $\pm\phn1.6\%$ & \nodata & $\pm2.6\%$ & $\pm1.0\%$ & $\pm\phn3.2\%$ & 2.18 \\
\nodata & \Neeight\tablenotemark{c} & $>2.35\e{-13}$ & \nodata & \nodata & \nodata & \nodata & \nodata & \nodata \\
HS 1102+3441 & \Osix & $1.85\e{-13}$ & $\pm\phn0.5\%$ & $\pm\phn7.5\%$ & $\pm2.6\%$ & $\pm0.5\%$ & $\pm\phn8.0\%$ & 62.63 \\
\nodata & \Nthree & $7.50\e{-14}$ & $\pm\phn0.9\%$ & $\pm15.0\%$ & $\pm2.7\%$ & $\pm0.7\%$ & $\pm15.3\%$ & 24.53\\
\nodata & \Cthree & $3.26\e{-14}$ & $\pm\phn2.3\%$ & $\pm15.0\%$ & $\pm3.7\%$ & $\pm1.0\%$ & $\pm15.6\%$ & 10.54 \\
\nodata & \Otwo + \Othree & $1.15\e{-14}$ & $\pm\phn4.9\%$ & \nodata & $\pm3.6\%$ & $\pm1.4\%$ & $\pm\phn6.2\%$ & 3.28 \\
\nodata & \Neeight & $7.25\e{-14}$ & $\pm\phn0.8\%$ & \nodata & $\pm4.6\%$ & $\pm2.0\%$ & $\pm\phn5.1\%$ & 19.47 \\
SDSS J124154.02+572107.3 & \Osix & $2.28\e{-14}$ & $\pm\phn2.7\%$ & $\pm\phn7.5\%$ & $\pm0.8\%$ & $\pm0.1\%$ & $\pm\phn8.0\%$ & 14.40\\
\nodata & \Nthree & $5.26\e{-15}$ & $\pm\phn6.2\%$ & $\pm15.0\%$ & $\pm0.7\%$ & $\pm0.1\%$ & $\pm16.2\%$ & 3.13\\
\nodata & \Cthree & $5.93\e{-15}$ & $\pm\phn4.9\%$ & $\pm15.0\%$ & $\pm0.6\%$ & $\pm0.1\%$ & $\pm15.8\%$ & 3.45\\
\nodata & \Neeight & $2.73\e{-14}$ & $\pm\phn1.6\%$ & \nodata & $\pm1.1\%$ & $\pm0.5\%$ & $\pm\phn2.1\%$ & 11.47 \\
HE 0238-1904 & \Osix & $3.13\e{-13}$ & $\pm\phn3.5\%$ & $\pm\phn7.5\%$ & $\pm2.7\%$ & $\pm0.3\%$ & $\pm\phn8.7\%$ & 22.07 \\
\nodata & \Nthree & $8.23\e{-14}$ & $\pm13.5\%$ & $\pm15.0\%$ & $\pm1.7\%$ & $\pm0.2\%$ & $\pm20.2\%$ & 5.78\\
\nodata & \Cthree & $3.36\e{-14}$ & $\pm\phn3.3\%$ & $\pm15.0\%$ & $\pm0.7\%$ & $\pm0.2\%$ & $\pm15.4\%$ & 2.36\\
\nodata & \Otwo + \Othree & $2.50\e{-14}$ & $\pm\phn5.8\%$ & \nodata & $\pm1.8\%$ & $\pm0.9\%$ & $\pm\phn6.2\%$ & 1.73 \\
\nodata & \Neeight & $1.15\e{-13}$ & $\pm\phn2.9\%$ & \nodata & $\pm3.4\%$ & $\pm1.5\%$ & $\pm\phn4.7\%$ & 7.88 \\
RXJ 2154.1-4414 & \Nfive & $6.50\e{-14}$ & $\pm\phn1.8\%$ & $\pm20.0\%$ & $\pm1.4\%$ & $\pm0.2\%$ & $\pm20.1\%$ & 9.23 \\
\nodata & Ly$\alpha$ & $9.59\e{-13}$ & $\pm\phn0.3\%$ & $\pm\phn2.0\%$ & $\pm1.4\%$ & $\pm0.2\%$ & $\pm\phn2.4\%$ & 132.63 \\
\nodata & \Osix & $2.36\e{-13}$ & $\pm\phn0.3\%$ & $\pm\phn7.5\%$ & $\pm1.3\%$ & $\pm0.5\%$ & $\pm\phn7.6\%$ & 26.36 \\
\nodata & \Nthree & $3.05\e{-14}$ & $\pm\phn1.6\%$ & $\pm15.0\%$ & $\pm0.9\%$ & $\pm0.4\%$ & $\pm15.1\%$ & 3.22 \\
\nodata & \Cthree & $2.63\e{-14}$ & $\pm\phn2.0\%$ & $\pm15.0\%$ & $\pm0.8\%$ & $\pm0.5\%$ & $\pm15.2\%$ & 2.72 \\
SDSS J132222.68+464535.2 & \Nfive & $1.06\e{-14}$ & $\pm\phn7.7\%$ & $\pm20.0\%$ & $\pm1.7\%$ & $\pm0.2\%$ & $\pm21.5\%$ & 5.60 \\
\nodata & Ly$\alpha$ & $2.33\e{-13}$ & $\pm\phn0.7\%$ & $\pm\phn2.0\%$ & $\pm1.7\%$ & $\pm0.2\%$ & $\pm\phn2.7\%$ & 120.58\\
\nodata & \Osix & $6.11\e{-14}$ & $\pm\phn1.0\%$ & $\pm\phn7.5\%$ & $\pm1.6\%$ & $\pm0.5\%$ & $\pm\phn7.7\%$ & 25.37 \\
\nodata & \Nthree & $8.30\e{-15}$ & $\pm\phn4.6\%$ & $\pm15.0\%$ & $\pm1.6\%$ & $\pm0.6\%$ & $\pm15.8\%$ & 3.25 \\
\nodata & \Cthree & $8.03\e{-15}$ & $\pm\phn3.8\%$ & $\pm15.0\%$ & $\pm1.5\%$ & $\pm0.6\%$ & $\pm15.6\%$ & 3.09 \\
\enddata
\tablenotetext{a}{Fluxes are in units of erg s$^{-1}$ cm$^{-2}$}
\tablenotetext{b}{All errors given as percentages of the measured flux}
\tablenotetext{c}{\,Lower limit determined by direct integration}
\end{deluxetable*}
\end{turnpage}
\clearpage
\global\pdfpageattr\expandafter{\the\pdfpageattr/Rotate 90}

\end{document}